\documentclass[twocolumn,showpacs,preprintnumbers,amsmath,amssymb]{revtex4}

\usepackage{graphicx}
\usepackage{dcolumn}
\usepackage{bm}
\usepackage{epsf}

\newcommand{\rmj}{{\rm j}}

\newcommand{\bea}{\begin{eqnarray}}

\newcommand{\eea}{\end{eqnarray}}

\begin{document}

\bibstyle{plain}


\title{Sampling Distributions of Random Electromagnetic Fields in\\ Mesoscopic or Dynamical Systems
}

\author{{\it Luk R. Arnaut}\\
\normalsize Time, Quantum and Electromagnetics Division\\
\normalsize National Physical Laboratory,
\normalsize Teddington,
\normalsize United Kingdom\\
\small and\\
\normalsize Department of Electrical and Electronic Engineering\\ 
\normalsize Imperial College of Science, Technology and Medicine\\ 
\normalsize South Kensington Campus, London, United Kingdom
}%

\date{\today}

\begin{abstract}
We derive the sampling probability density function (pdf) of an ideal localized random electromagnetic field, its amplitude and intensity in an  electromagnetic environment that is quasi-statically time-varying statistically homogeneous or static statistically inhomogeneous. The results allow for the estimation of field statistics and confidence intervals when a single spatial or temporal stochastic process produces randomization of the field. Sampling distributions are particularly significant when the number of degrees of freedom $\nu$ is relatively small (typically, $\nu<40$), e.g., in mesoscopic systems when the sample set size $N$ is relatively small by choice or by force. Results for both coherent and incoherent detection techniques are derived, for Cartesian, planar and full-vectorial fields. We show that the functional form of the sampling pdf depends on whether the random variable is dimensioned (e.g., the sampled electric field proper) or is expressed in dimensionless standardized or normalized form (e.g., the sampled electric field divided by its sampled standard deviation). For dimensioned quantities, the electric field, its amplitude and intensity exhibit different types of Bessel $K$ sampling pdfs, which differ significantly from the asymptotic Gauss normal and $\chi^{(2)}_{2p}$ ensemble pdfs when $\nu$ is relatively small. By contrast, for the corresponding standardized quantities, Student $t$, Fisher-Snedecor $F$ and root-$F$ sampling pdfs are obtained that exhibit heavier tails than comparable Bessel $K$ pdfs. Statistical uncertainties obtained from classical small-sample theory for dimensionless quantities are shown to be overestimated compared to dimensioned quantities. Differences in the sampling pdfs arising from de-normalization versus de-standardization are obtained.
\end{abstract}

                              
\maketitle

                                                            
\section{Introduction \label{sec:intro}}
In the application of statistical methods to disordered and chaotic systems, particularly nonergodic and/or mesoscopic systems \cite{scha1}--\cite{chab1}, the role played by sampling distributions \cite{cram1} is of fundamental importance because, in practice, data sets are necessarily of limited (and often small) size $N$. Strictly, the Central Limit Theorem (CLT) is inapplicable to finite sample sets, {\it a fortiori} to small sets.
Therefore, sampling distributions, rather than their underlying (parent) ensemble distributions, should always be used in any proper comparison between theoretical and empirical probability distributions based on numerical data from practical measurement (experiment) or simulation (computation), particularly when detecting or validating unexpected phenomena.
The finiteness of the number of degrees of freedom for sample sets, $\nu$, can have a profound effect not only on the statistical moments, but also on the functional form and shape of the probability density function (pdf) -- particularly near its extremities -- and on second- or higher-order statistics, such as auto- or cross-correlation functions and associated spectral densities, etc. 
As a matter of fact, sampling pdfs draw on properties of both first- and second-order ensemble statistics: while the sampling distributions are representations for local instantaneous sampled fields, the value of $\nu$ as a sampling distribution parameter is governed by the correlation distance between points. This dependence on nonlocal properties of the field applies to any physically limited realizable sampled region.
For sample sets in which all $N$ values are statistically independent (as we shall further assume throughout), $\nu=N-1$.

Ensemble pdfs for ideal random 3-D electromagnetic (EM) fields are well known and have been amply investigated in various physical applications, whether in unbounded space \cite{kost1} or in the presence of an impedance boundary \cite{arnaRS}. By contrast, their associated sampling pdfs have received little attention yet are pertinent and require characterization.
Idealized random (stochastic) classical EM fields offer a paradigm 
for characterizing wave propagation or transport governed by dynamic multiple scattering (time-varying configuration or boundaries). 
Examples include 
fields inside acoustic or EM mode-tuned or mode-stirred reverberation chambers (MT/MSRCs) \cite{kost1}--\cite{rosn2};
mesoscopic structures;
random or turbulent media;
polarization and anisotropy in, e.g., the cosmic microwave background radiation \cite{kami1}--\cite{rath1};
static or dynamic optical diffusion using random phase screens; 
diffusing wave spectroscopy \cite{mart1}--\cite{pear1}; etc.

For the estimation of statistics of the field, amplitude, intensity, energy density and power for such `wave chaos', 
the ensemble pdf of the field is usually adequate when the number of statistically independent partial contributions and, hence, $\nu$ approaches infinity. 
For example, for MT/MSRCs, the prerequisites are that the wavelength is very short relative to the physical dimensions of the cavity and that the observation time is long compared to the correlation time of the stirring process because, physically, a random EM field is governed by the spatial or temporal extent of the space- or time-limited process. 
On occasion, however, $\nu$ is too small for these ensemble pdfs to be sufficiently accurate. 
In this case, the CLT or maximum-entropy principle may not be applicable. 
Nevertheless, in such low-dimensional cases, spatial and/or temporal averaging increases the value of $\nu$ \cite{arnalocavg}, which may have a substantial effect on the measured or perceived distribution of the field.

A key issue addressed in this paper is that statistically homogeneous random fields are characterized by pdfs with distribution parameters (e.g., average, standard deviation, number of degrees of freedom, etc.) whose values have to be {\em estimated\/} from the {\em same\/} set of data of the fluctuating field. As a result, these parameters themselves show sampling fluctuations because of the finite value of $\nu$. These variations give rise to bi- or multivariate fluctuations and increased uncertainty (i.e., wider confidence intervals) for the sampled field, compared to the corresponding ensemble distribution where these parameters are known exactly. 

The physical origin of the limitation of the value of $\nu$ can be twofold.
First, the potential (i.e., maximum attainable) number of {\em statistically independent\/} realizations, $N_{\rm max}$, may be practically restricted, even if an unlimited number of different states of the statistical ensemble were physically realized. For example, in a MT/MSRC, this case corresponds to the so-called undermoded regime (i.e., multi-mode operation but with less than, say, ten cavity modes being excited simultaneously); typically occurring at wavelengths where modal overlap is relatively small.
As another example, in a relatively small sample of a random medium, the limitation on the value of $\nu$ refers to the case of a relatively small loading fraction of inclusions. In this case, even the ensemble distribution does not possess Gauss normal statistics.
Secondly, $N_{\rm max}$ may be unlimited but, for economical or other reasons, the sample size may have to be severely restricted ($N\ll N_{\rm max}$).
In this case, while $\nu_{\rm max}$ is potentially large, \rm the value of $\nu$ actually realized in a sample set is relatively small.

In this paper, we investigate sampling pdfs of the complex value, amplitude and intensity of 1-D, 2-D and 3-D random EM fields that have ideal Gaussian ensemble probability distributions, for both the actual (non-standardized) observed EM quantities and for their standardized forms.
Here, the notions of standarized and normalized random variables refer to a random field quantity divided by its own (sample or ensemble) standard deviation or mean value, respectively.
Coherent as well as incoherent detection techniques are considered in each case. However, specific nonstationary effects associated with the dynamics of multiple scattering, which are relevant to certain aspects of diffusing wave spectroscopy \cite{mare1,zakh1}, are not addressed here: each realized state of the system is considered in its quasi-stationary 
approximation.
 The scenario investigated here is also relevant, in particular, to several practical applications in wireless communications (e.g., 
mobile-to-fixed and mobile-to-mobile transmission), signal processing,
wave propagation in turbulent media, 
modal noise in optical fibres under restricted-mode launch conditions \cite{pape1},
etc.

\section{Electric or magnetic field\label{sec:field}}
\subsection{Coherent detection}
\subsubsection{Non-standardized field\label{sec:samplepdfE_coh}}
Consider a modulated local analytic electric field $E=E^\prime-\rmj E^{\prime\prime}$ received by an antenna (sensor) or scatterer immersed in a time-varying multi-scattering environment. A harmonic time dependence $\exp({\rm j} \omega t)$ is assumed and suppressed.
If this field is made up of an arbitrarily large (theoretically infinite) number of fluctuating partial fields (i.e., modal or angular plane-wave spectral components) whose realization forms a random walk in the complex plane \cite{arnaKdf}, then, on account of the CLT (valid under very general but definite conditions \cite{beck2}), the associated 
conditional probability 
of $E^{\prime(\prime)}$ coincides with the ensemble pdf, given by
\bea
f_{E^{\prime(\prime)} | (S_{E^{\prime(\prime)}},M_{E^{\prime(\prime)}})}(e^{\prime(\prime)}|(s_{E^{\prime(\prime)}},m_{E^{\prime(\prime)}}))
\nonumber\\
=
\frac{\exp \left [ - \frac{\left ( e^{\prime(\prime)} - m_{E^{\prime(\prime)}} \right )^2}{2 \thinspace s^2_{E^{\prime(\prime)}}} \right ] }{\sqrt{2\pi}~s_{E^{\prime(\prime)}}}.
\label{eq:cpdf}
\eea
Here, $M_{E^{\prime(\prime)}}$ and $S_{E^{\prime(\prime)}}$ represent random variables induced by the sample mean and sample standard deviation of $E^{\prime(\prime)}$, respectively, whereas lowercase symbols $m_{E^{\prime(\prime)}}$ and $s_{E^{\prime(\prime)}}$ represent their corresponding sample values.
If the mean and standard deviation of $E^{\prime(\prime)}$ are known with certainty, with respective constant ensemble values $\mu_{E^{\prime(\prime)}}$ and $\sigma_{E^{\prime(\prime)}}$, i.e., 
\bea
f_{M_{E^{\prime(\prime)}}}(m_{E^{\prime(\prime)}}) 
&=& \delta(m_{E^{\prime(\prime)}}-\mu_{E^{\prime(\prime)}})\label{eq:deltapdfmu}\\
f_{S_{E^{\prime(\prime)}}}(s_{E^{\prime(\prime)}}) 
&=& \delta(s_{E^{\prime(\prime)}}-\sigma_{E^{\prime(\prime)}})\label{eq:deltapdfsigma}
\eea
then the conditional pdf (\ref{eq:cpdf}) coincides with the marginal pdf $f_{E^{\prime(\prime)}}({e^{\prime(\prime)}})$ because, in general,
\bea
&~& \hspace{-0.8cm}
f_{E^{\prime(\prime)}}({e^{\prime(\prime)}}) \nonumber\\
&=&
\int^{+\infty}_{-\infty} {\rm d} m_{E^{\prime(\prime)}}
\int^{+\infty}_{0} {\rm d} s_{E^{\prime(\prime)}} 
\nonumber\\ &~&\times
f_{E^{\prime(\prime)}, S_{E^{\prime(\prime)}},M_{E^{\prime(\prime)}}}(e^{\prime(\prime)},s_{E^{\prime(\prime)}},m_{E^{\prime(\prime)}})\\
&=&
\int^{+\infty}_{-\infty} {\rm d} m_{E^{\prime(\prime)}}
\int^{+\infty}_{0} {\rm d} s_{E^{\prime(\prime)}}
\nonumber\\ &~&\times
f_{E^{\prime(\prime)} | (S_{E^{\prime(\prime)}},M_{E^{\prime(\prime)}})}(e^{\prime(\prime)}|(s_{E^{\prime(\prime)}},m_{E^{\prime(\prime)}}))
\nonumber\\
&~&\times
f_{S_{E^{\prime(\prime)}},M_{E^{\prime(\prime)}}}(s_{E^{\prime(\prime)}},m_{E^{\prime(\prime)}})
\label{eq:jpdf_gen}
\eea
and because in (and only in) the case of a Gauss normal ensemble distribution of $E^{\prime(\prime)}$ are the sample mean and sample standard deviation independent random variables \cite{cram1}, i.e.,
\bea
f_{S_{E^{\prime(\prime)}},M_{E^{\prime(\prime)}}}(s_{E^{\prime(\prime)}},m_{E^{\prime(\prime)}})
=
f_{S_{E^{\prime(\prime)}}}(s_{E^{\prime(\prime)}})
f_{M_{E^{\prime(\prime)}}}(m_{E^{\prime(\prime)}}).
\nonumber\\
\label{eq:jpdf_indep}
\eea
From (\ref{eq:cpdf}) and (\ref{eq:deltapdfmu})--(\ref{eq:jpdf_indep}), we obtain
\bea
f_{E^{\prime(\prime)}}(e^{\prime(\prime)})
=
\frac{\exp \left [ - \frac{\left ( e^{\prime(\prime)} - \mu_{E^{\prime(\prime)}} \right )^2}{2 \thinspace \sigma^2_{E^{\prime(\prime)}}} \right ] }{\sqrt{2\pi}~\sigma_{E^{\prime(\prime)}}}.
\label{eq:mpdf}
\eea
Here, $\sigma_{E^{\prime(\prime)}} = \sigma_{E}/\sqrt{2} = \sqrt{p}\thinspace \sigma_{E^{(\prime)\prime}_\alpha} = \sqrt{p/2}\thinspace \sigma_{E_\alpha}$ where $\alpha=x,y$ or $z$ for the three Cartesian components $E_\alpha$ of ${\bf E}$ and where the value of $p$ corresponds to the number of spatial dimensions in which ${\bf E}$ is being considered. 

Of the possible values of $p$, viz., $1$, $2$, or $3$, its specific value is governed by the polarization direction of the electric-field sensor and/or by any EM excitation or boundary conditions that may enforce a certain fixed (deterministic) polarization state of the field.
Thus, the Cartesian component of a vector field and the 3-D vector field itself correspond to the cases $p=1$ and $p=3$, respectively. For a paraxially propagating polarized or unpolarized optical random field, $p=1$ or $p=2$, respectively. The case $p=2$ refers to an unpolarized electric field that is transverse to the local wavevector (i.e., randomly elliptically polarized field) and will be denoted by a subscript `t', whereas $p=1$ corresponds to its polarized detection or to a randomly modulated linearly polarized field, denoted by a subscript `$\alpha$'.
These values of $p$ apply to local fields detected by an electrically small sensor, whose characteristic length is less than the spatial coherence length of the field, i.e., typically $\lambda/2$ in an unbounded medium. Larger values of $p$ apply to electrically large detectors whose receiving cross-section (whether physical or as a synthetic aperture) is larger than $\lambda/2$ in one or more dimensions. For example, for currents induced in a linear antenna of length $L$, we have $p={\rm max} [1, L/(\lambda/2)]$.

If, instead of (\ref{eq:deltapdfsigma}), $S_{E^{\prime(\prime)}}$ exhibits random fluctuations or if its value is not known precisely, the sampling pdf of $S_{E^{\prime(\prime)}}$ for Gaussian $E^{\prime(\prime)}|S_{E^{\prime(\prime)}}$ is a $\chi_{pN-1}$ pdf, i.e.,
\bea f_{S_{E^{\prime(\prime)}}}(s_{E^{\prime(\prime)}};N) &=& 
\frac{C_{S_{E^{\prime(\prime)}}}}{\sigma_{S_{E^{\prime(\prime)}}}} 
\left ( \frac{s_{E^{\prime(\prime)}}}{\sigma_{S_{E^{\prime(\prime)}}}} \right )^{pN-2}
\nonumber\\&~& \times
\exp \left [ - {\cal N} \left ( \frac{s_{E^{\prime(\prime)}}}{\sigma_{S_{E^{\prime(\prime)}}}} \right )^2 \right ]
\label{eq:pdfStdX_Gauss}
\eea
with
\bea
C_{S_{E^{\prime(\prime)}}} &\stackrel{\Delta}{=}&
\frac{2~{\cal N}^{\frac{pN-1}{2}}}{\Gamma\left ( \frac{pN-1}{2} \right )},\\
{\cal N} &\stackrel{\Delta}{=}& {\frac{pN-1}{2} - \left [ \frac{\Gamma\left ( \frac{pN}{2} \right )}{\Gamma \left ( \frac{pN-1}{2} \right )}\right ]^2}.
\eea
The sampling pdf (\ref{eq:pdfStdX_Gauss}) is in its self-sufficient form \cite{arnalocavg}, i.e., it contains the standard deviation of $S_{E^{\prime(\prime)}}$ itself as a distribution parameter. Alternatively, (\ref{eq:pdfStdX_Gauss}) can be re-expressed in terms of the ensemble standard deviation $\sigma_{E^{\prime(\prime)}}$ of $E^{\prime(\prime)}$, because both statistics are related via
\bea
\sigma_{S_{E^{\prime(\prime)}}} &=& \sigma_{E^{\prime(\prime)}} \sqrt{\frac{2~{\cal N}}{pN-1}} .
\label{eq:stdstdX}
\eea
With the aid of \cite[(3.471.9)]{grad1}, the sampling pdf of $E^{\prime(\prime)}$ for $\mu_{E^{\prime(\prime)}}=0$ in (\ref{eq:jpdf_gen}) is obtained as a marginal pdf of the joint pdf $f_{E^{\prime(\prime)},S_{E^{\prime(\prime)}}}(e^{\prime(\prime)},s_{E^{\prime(\prime)}};N)$, viz.,
\bea
&~&\hspace{-0.8cm} f_{E^{\prime(\prime)}}(e^{\prime(\prime)};N)\nonumber\\ 
&=& \int^{+\infty}_0 f_{E^{\prime(\prime)}|S_{E^{\prime(\prime)}}}(e^{\prime(\prime)}|s_{E^{\prime(\prime)}}) f_{S_{E^{\prime(\prime)}}}(s_{E^{\prime(\prime)}};N) {\rm d}s_{E^{\prime(\prime)}} \nonumber\\
&=& \frac{(pN-1)^{\frac{pN-1}{2}}}{2^{\frac{pN}{2}} \sqrt{\pi}\thinspace \Gamma \left ( \frac{pN-1}{2} \right ) ~\sigma^{pN-1}_{E^{\prime(\prime)}}} 
\nonumber\\&~&\times
\int^{+\infty}_0 t^{\frac{pN}{2}-2} \exp \left ( - \frac{e^{{\prime(\prime)}^2}}{2t} - \frac{pN-1}{2} \frac{t}{\sigma^2_{E^{\prime(\prime)}}} \right ) {\rm d} t,
\nonumber\\
\eea
i.e.,
\bea
&~& \hspace{-1.3cm} f_{E^{\prime(\prime)}}(e^{\prime(\prime)};N)
=\nonumber\\
&~& \hspace{-0.8cm} \frac{C_{E^{\prime(\prime)}}}{\sigma_{E^{\prime(\prime)}}} \left ( \frac{|e^{\prime(\prime)}|}{\sigma_{E^{\prime(\prime)}}} \right )^{\frac{pN}{2}-1} K_{\frac{pN}{2}-1} \left ( \sqrt{pN-1} \frac{|e^{\prime(\prime)}|}{\sigma_{E^{\prime(\prime)}}} \right )
\label{eq:pdfE_BesselK}
\label{eq:samplepdfE_coh_BesselK}
\eea
with
\bea
C_{E^{\prime(\prime)}} \stackrel{\Delta}{=} \frac{(pN-1)^{\frac{pN}{4}}}{2^{\frac{pN}{2}} \sqrt{\pi} \thinspace \Gamma \left ( \frac{pN-1}{2} \right ) }.
\label{eq:def_CE}
\eea
The pdf (\ref{eq:pdfE_BesselK}) belongs to McKay's class of Bessel $K$ distributions \cite{mckay1}. These pdfs have also been obtained in a variety of applied statistical problems, in electromagnetics and elsewhere; cf., e.g., \cite{nolt1}, \cite{naka1} and references in \cite{arnaKdf}.
It is further worth noting that the pdf (\ref{eq:pdfE_BesselK}) has also been obtained in a different case, viz., as an ensemble pdf associated with an {\it ad hoc} Bessel $K$ distributed field amplitude (implying non-Gaussian/non-Rayleigh ensemble statistics) and derived via a Blanc-Lapierre transformation \cite{prim1}.
Results that are at least qualitatively compliant with those obtained from the pdf (\ref{eq:pdfE_BesselK}) have been observed recently in experiments involving correlated scattering \cite{chab1}. The pdf (\ref{eq:pdfE_BesselK}) is also included in the class of Meijer $G$ limit pdfs for complex fields in undermoded MT/MSRCs \cite{arnaKdf}, which in turn is a special case of more general Fox $H$ distributions \cite{sprin1}.

It is emphasized that the result (\ref{eq:pdfE_BesselK}) strictly holds for a physically ideal, viz., Gaussian random field if it is subjected to a finite-sized sampling (detection) process. Only in this case is a Gaussian marginal pdf $f_{E^{\prime(\prime)}|S_{E^{\prime(\prime)}}}$ compatible with a non-delta-distributed ${S_{E^{\prime(\prime)}}}$. For the physical field itself, a Gaussian $f_{E^{\prime(\prime)}|S_{E^{\prime(\prime)}}}$ implies the number of degrees of freedom approaching infinity, hence resulting in a delta distribution for $f_{S_{E^{\prime(\prime)}}}$, and vice versa.

Fig. \ref{fig:pdfE_paramNt}a shows the pdf (\ref{eq:samplepdfE_coh_BesselK}) for selected values of $N$. Because of symmetry of $f_{E^{\prime(\prime)}}$ with respect to its mean value, only the positive half of the distribution is shown.

If $E^{\prime(\prime)}$ exhibits a deterministic, i.e., constant bias $\mu_{E^{\prime(\prime)}}$ as in (\ref{eq:mpdf}), then 
\bea
f_{E^{\prime(\prime)}}(e^{\prime(\prime)};N) 
&=&
\frac{C_{E^{\prime(\prime)}}}{\sigma_{E^{\prime(\prime)}}} \left ( \frac{|e^{\prime(\prime)}-\mu_{E^{\prime(\prime)}}|}{\sigma_{E^{\prime(\prime)}}} \right )^{\frac{pN}{2}-1} \nonumber\\
&~& \times K_{\frac{pN}{2}-1} \left ( \sqrt{pN-1} \frac{|e^{\prime(\prime)}-\mu_{E^{\prime(\prime)}}|}{\sigma_{E^{\prime(\prime)}}} \right ).
\nonumber\\
\label{eq:pdfE_BesselK_determbias}
\label{eq:samplepdfE_coh_BesselK_determbias}
\eea

For nonnegligible ``slow'' fluctuations of $m_{E^{\prime(\prime)}}$ in (\ref{eq:cpdf}), i.e., for a random bias (trend) of $E^{\prime(\prime)}$, the results are easily generalized as follows. 
On account of linearity, the sampling variable $M_{E^{\prime(\prime)}}$ for Gaussian $E^{\prime(\prime)}$ is also Gaussian with the same expected value but standard deviation $\sigma_{E^{\prime(\prime)}}/\sqrt{N}$. 
Therefore, if the bias is itself random with Gauss normal distribution, i.e.,
\bea
f_{M_{E^{\prime(\prime)}}}(m_{E^{\prime(\prime)}};N)
=
\frac{\exp \left [ - \frac{N\left ( m_{E^{\prime(\prime)}} - \mu_{E^{\prime(\prime)}} \right )^2}{2 \thinspace \sigma^2_{E^{\prime(\prime)}}} \right ] }{\sqrt{2\pi/N}~\sigma_{E^{\prime(\prime)}}}
\label{eq:mpdf2}
\eea
instead of (\ref{eq:deltapdfmu}), then with (\ref{eq:jpdf_gen})--(\ref{eq:jpdf_indep}) we arrive at
\bea
&~& \hspace{-1cm}
f_{E^{\prime(\prime)}}(e^{\prime(\prime)};N) =\nonumber\\
&~& \hspace{-0.5cm} \frac{C^\prime_{E^{\prime(\prime)}}}{\sigma_{E^{\prime(\prime)}}} 
\int^{+\infty}_{-\infty}
\left |e^{\prime(\prime)}-m_{E^{\prime(\prime)}} \right |^{\frac{pN}{2}-1} \nonumber\\ 
&~& \hspace{-0.5cm} \times
\exp \left [ - \frac{N\left ( m_{E^{\prime(\prime)}} - \mu_{E^{\prime(\prime)}} \right )^2}{2 \thinspace \sigma^2_{E^{\prime(\prime)}}} \right ] \nonumber\\
&~& \hspace{-0.5cm} \times K_{\frac{pN}{2}-1} \left ( \sqrt{pN-1} \frac{|e^{\prime(\prime)}-m_{E^{\prime(\prime)}}|}{\sigma_{E^{\prime(\prime)}}} \right ) {\rm d} m_{E^{\prime(\prime)}}.
\label{eq:pdfE_BesselK_randombias}
\label{eq:samplepdfE_coh_BesselK_randombias}
\eea
For practical numerical integration of (\ref{eq:pdfE_BesselK_randombias}), the double-infinite range of $m_{E^{\prime(\prime)}}$ can be limited to, say, 
$[\mu_{E^{\prime(\prime)}}-5 \sigma_{E^{\prime(\prime)}}/\sqrt{N},\mu_{E^{\prime(\prime)}}+5 \sigma_{E^{\prime(\prime)}}/\sqrt{N}]$.

For sufficiently large $N$, the field $E^{\prime(\prime)}$ and its sample mean $M_{E^{\prime(\prime)}}$ are approximately independent (sharing $N-1$ degrees of freedom), whence $E^{\prime(\prime)} - M_{E^{\prime(\prime)}}$ is then also Gaussian with zero mean and standard deviation $\sigma^*_{E^{\prime(\prime)}}\simeq\sigma_{E^{\prime(\prime)}}\sqrt{1+1/N}$.
Thus, (\ref{eq:pdfE_BesselK_determbias}) with $\sigma_{E^{\prime(\prime)}}$ replaced by $\sigma^*_{E^{\prime(\prime)}}$ serves as a first-order approximation to the pdf (\ref{eq:pdfE_BesselK_randombias}) for large $N$.

\subsubsection{Standardized field}
For comparison of an experimental distribution against a theoretical sampling distribution, it is necessary to standardize the experimentally obtained $E^{\prime(\prime)}$. Then, rather than determining the pdf of $E^{\prime(\prime)}$, we require the pdf of the dimensionless variate
\bea
X^{\prime(\prime)} \stackrel{\Delta}{=} \frac{E^{\prime(\prime)} - M_{E^{\prime(\prime)}}}{S_{E^{\prime(\prime)}}}
\label{eq:def_StandarizedE}
\eea
i.e.,
for uncertain sample values of $M_{E^{\prime(\prime)}}$ and $S_{E^{\prime(\prime)}}$ that are unknown {\it a priori}. 
Their values are often to be estimated from the same limited sample data set of $E^{\prime(\prime)}$.
For Gaussian $E^{\prime(\prime)}$, the $M_{E^{\prime(\prime)}}$ and $S_{E^{\prime(\prime)}}$ are statistically independent and their sampling distributions are Gaussian and $\chi_{pN-1}$, respectively \cite{arnaTQE2}.
We further assume that the fluctuations of $M_{E^{\prime(\prime)}}$ are small compared to those of $S_{E^{\prime(\prime)}}$ and, {\it a fortiori}, $E^{\prime(\prime)}$. 

It is shown in Sec. \ref{app:samplepdfX} that $X^{\prime(\prime)}$ for $M_{E^{\prime(\prime)}}=0$ exhibits a Student $t$ distribution \cite{stud1} with $pN-1$ degrees of freedom, i.e.,
\bea
f_{X^{\prime(\prime)}}(x^{\prime(\prime)};N) = 
C_{X^{\prime(\prime)}}
\left ( 1 + \frac{{x^{\prime(\prime)}}^2}{pN-1} \right )^{-{pN}/{2}}
\label{eq:pdfE_temp}
\label{eq:samplepdfX_coh_StudentT}
\eea
with $X^{\prime(\prime)} \stackrel{\Delta}{=} E^{\prime(\prime)}/S_{E^{\prime(\prime)}}$ for $\langle E^{\prime(\prime)} \rangle=0$, where the sample value $x^{\prime(\prime)} = e^{\prime(\prime)}/s_{E^{\prime(\prime)}}$ is estimated as a (dimensionless) ratio of the two sample values $e^{\prime(\prime)}$ and $s_{E^{\prime(\prime)}}$, and
where
\bea
C_{X^{\prime(\prime)}} \stackrel{\Delta}{=} \frac{\Gamma \left ( \frac{pN}{2} \right )}{\Gamma \left ( \frac{pN-1}{2} \right )\sqrt{(pN-1)\pi} }.
\eea
The pdf (\ref{eq:samplepdfX_coh_StudentT}) can be used to compare the empirical distribution of the measured standardized complex field against its ideal theoretical sampling distribution, particularly when $N$ is relatively small ($N\sim 40$ or less).
For multi-scattering contributions, the value of $N$ can be associated with the number of scattering centers surrounding the receiver and/or the states during the motion of the environment around a momentarily stationary receiver.

\begin{figure}[htb] \begin{center} \begin{tabular}{cc} \ 
\epsfxsize=7cm \epsfbox{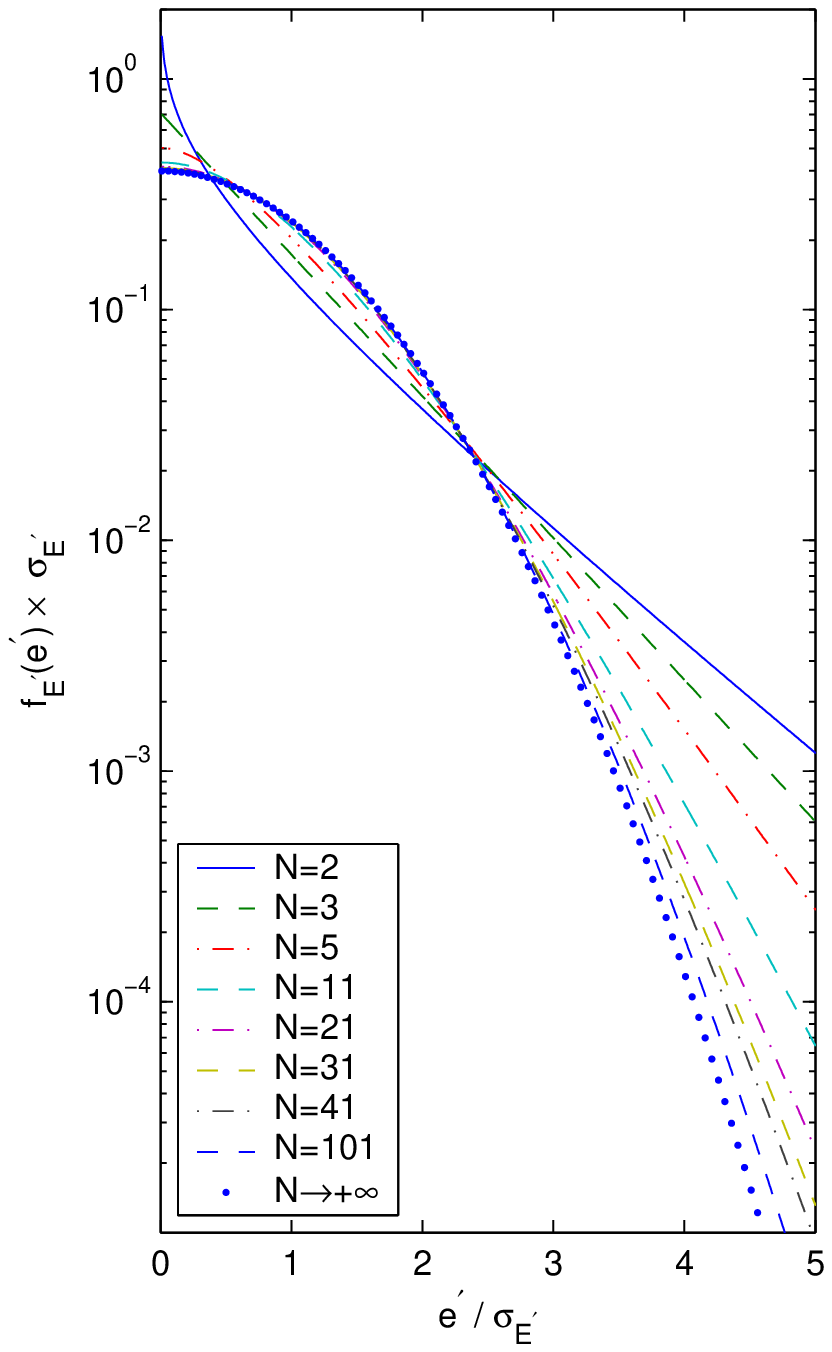}\ & \epsfxsize=7cm \epsfbox{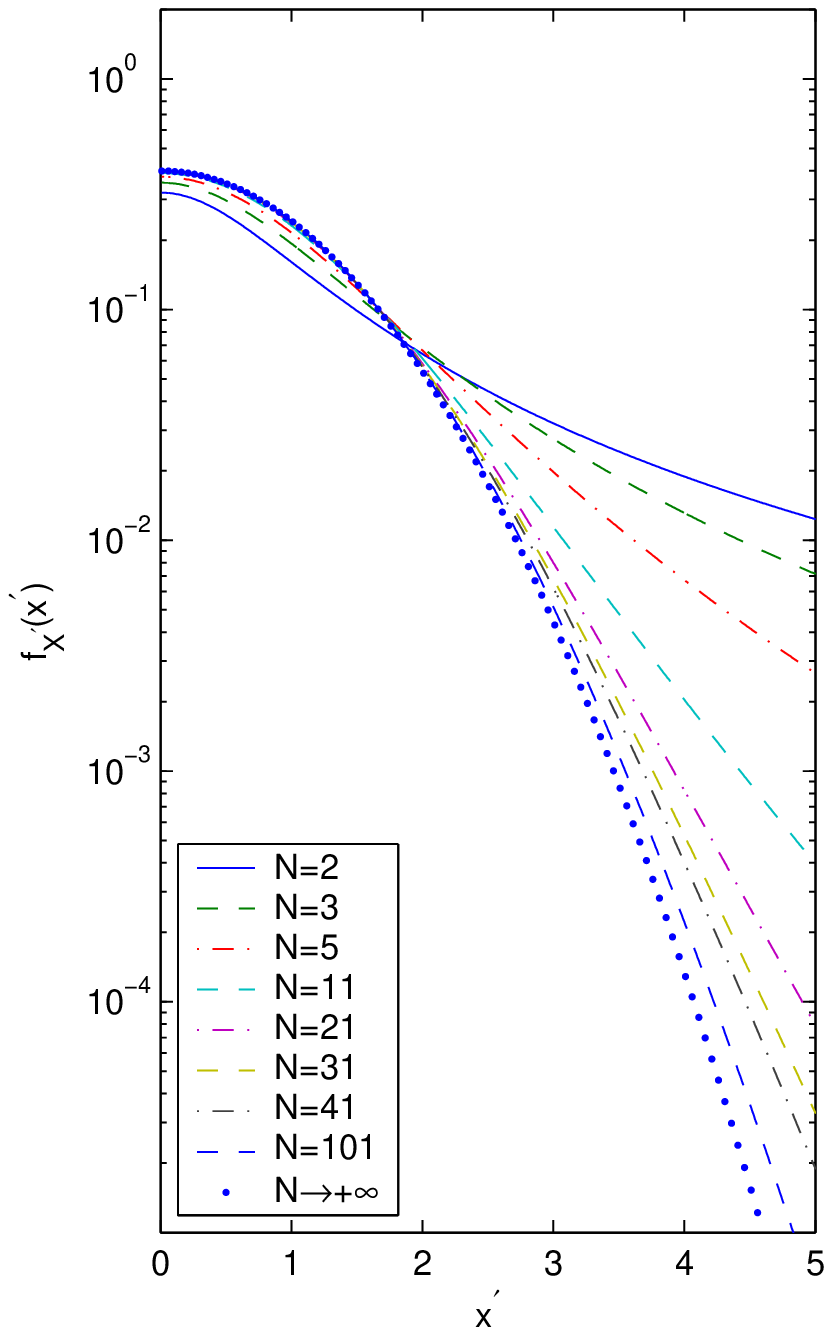}\ \\
\end{tabular} \end{center}
\caption{\label{fig:pdfE_paramNt}
\small
(color on-line)
Sampling probability density function for real or imaginary Cartesian component of electric field ($p{=}1$) at selected values of $N$: 
(a) Bessel $K$ sampling pdfs of $E^{\prime(\prime)}_\alpha$
[eq. (\ref{eq:samplepdfE_coh_BesselK})]; 
(b) Student $t$ sampling pdfs  of $X^{\prime(\prime)} = E^{\prime(\prime)}_\alpha / S_{{E^{\prime(\prime)}_\alpha}}$
[eq. (\ref{eq:samplepdfX_coh_StudentT})].
In both plots, the right tail becomes thinner for increasing values of $N$.
}
\end{figure}

Without loss of generality, we further consider a deterministically unbiased field ($\mu_{E^{\prime(\prime)}}=0$).
Fig. \ref{fig:pdfE_paramNt}b shows the pdf of $X^{\prime(\prime)}$ for selected values of $N$. Again, only the positive half of the distribution is shown.
It can be seen that, compared to the asymptotic Gauss normal pdf,
differences with $f_{E^{\prime(\prime)}}$ in Fig. \ref{fig:pdfE_paramNt}a occur mainly near the origin and in the tails.

\subsection{Incoherent detection}
Of significant practical importance, particular for measurements at optical wavelengths, is the case where the random field is sampled incoherently, i.e., by square-law detection of the electric or magnetic power, energy or intensity devoid of phase information. Yet one may wish to extract the sampling pdf of the complex-valued electric or magnetic field from such a scalar measurement.
To this end, we borrow results derived in Sec. \ref{sec:power_incoh},
and restrict further analysis to the case of a circular $f_E(e)$. (For an elliptic $f_E(e)$, data for a second linearly independent quadratic form of $E^\prime$ and $E^{\prime\prime}$ are needed, e.g., ${E^{\prime}}^2 - {E^{{\prime\prime}}}^2$.) From the derivation in Sec. \ref{sec:power_incoh}, it follows that the sampling pdf of $U^{\prime(\prime)} \stackrel{\Delta}{=} {E^{{\prime(\prime)}}}^2$ is (\ref{eq:samplepdfU_final}) after replacing $p$ with $p/2$, because the ensemble pdf of $U^{\prime(\prime)}|S_{U^{\prime(\prime)}}$ is $\chi^2_p$
whereas $f_{U^{\prime(\prime)}}$ has a $\chi^2_{pN-1}$ sampling pdf. With the variate transformation $f_{E^{\prime(\prime)}}(e^{\prime(\prime)}) = 2 |e^{\prime(\prime)}| f_{U^{\prime(\prime)}}(u^{\prime(\prime)}=e^{{\prime(\prime)}^2})$, we arrive at 
\bea
f_{E^{\prime(\prime)}}(e^{\prime(\prime)};N)
&=&
C^{\prime\prime}_{E^{\prime(\prime)}}
\left ( \frac{|e^{\prime(\prime)}|}{\sqrt{\sigma_U}} \right )^{\frac{1}{2} [p(N+1)-3]}
\nonumber\\ 
&~& \hspace{-1cm} \times K_{\frac{1}{2}[p(N-1)-1]}\left ( \sqrt{\sqrt{{2p}} \left ( {p} N - 1 \right )} \frac{|e^{\prime(\prime)}|}{\sqrt{\sigma_U}} \right ).
\nonumber\\
\label{eq:samplepdfE_incoh}
\eea
For detection of a Cartesian component of intensity or power, i.e., $p=1$, (\ref{eq:samplepdfE_incoh}) is equivalent with (\ref{eq:samplepdfE_coh_BesselK}).

\section{Field intensity (energy density, power)\label{sec:intensity}}
For the field intensity $U=U^\prime+U^{\prime\prime}=|E|^2$ -- as well as for the energy density or power, which are proportional to $U$ -- we can determine its pdf either on the basis of measured in-phase and/or quadrature components of the field (i.e., coherent detection, e.g., using a vector network analyzer), or measured directly using a square-law detector (i.e., incoherent detection, e.g., using a power meter, spectrum analyzer, field probe, etc.).
We determine sampling pdfs for both cases.

\subsection{Incoherent detection\label{sec:power_incoh}}
\subsubsection{Non-standardized intensity\label{sec:power_incoh_nonstand}}
In many practical cases, the intensity or energy density is measured or perceived through a square-law detector or perception process. 
Compared to $f_{E^{\prime(\prime)}|S_{E^{\prime(\prime)}}}$ for coherent detection, the pertinent conditional pdf (cpdf) is now $f_{U|S_{U}}$. On account of the CLT, the sampling cpdf for $U|S_{U}$ is $\chi^2_{2p}$, i.e., 
\bea
f_{U|S_U}(u|s_U) = \frac{p^{p/2}}{\Gamma(p) ~ s_{U}} \left ( \frac{u}{s_U} \right )^{p-1} \exp \left ( - \sqrt{p} \frac{u}{s_{U}} \right ).
\label{eq:cpdf_incoh_approx}
\eea
The sampling pdf of $S_U$, associated with an underlying circular Gauss normal $E$, is a $\chi^2_{2pN-1}$ pdf, i.e.,
\bea
f_{S_U}(s_U; N)
&=&
\frac{\left( pN - \frac{1}{2} \right)^{\frac{1}{2}\left (pN-\frac{1}{2} \right )}}
{\Gamma\left ( pN-\frac{1}{2} \right )\thinspace \sigma_{S_U}}
\left ( \frac{s_U}{\sigma_{S_U}} \right )^{pN-\frac{3}{2}} \nonumber\\ &~& \times \exp \left ( - \sqrt{pN - \frac{1}{2}}~\frac{ s_U}{\sigma_{S_U}} \, \right )
\label{eq:pdf_stdP_exact}
\eea
with 
\bea
\sigma_{S_U} = \frac{\sigma_U }{ \sqrt{pN-\frac{1}{2}} }.
\eea
Performing a similar calculation as in Sec. \ref{sec:samplepdfE_coh} and again using \cite[(3.471.9)]{grad1} yields the sampling pdf of $U$ as
a marginal pdf of $f_{U,S_U}(u,s_U;N)$. A Bessel $K$ distribution is again obtained, but compared to (\ref{eq:pdfE_BesselK}) it is now of a different type (i.e., the exponent of the power and the order of the Bessel function differ by a different amount), viz.,
\bea
f_U(u;N) &=&
\int^{+\infty}_0 f_{U|S_U}(u|s_U) f_{S_U}(s_u;N) {\rm d}s_U\label{eq:pdfU_nonstand_gen}\\
&=& \frac{C_U}{\sigma_{U}} \left (  \frac{u}{\sigma_{U}} \right )^{\frac{1}{2} \left [ p(N+1) - \frac{5}{2}\right ]} \nonumber\\ &~& \times K_{p(N-1)-\frac{1}{2}} \left ( 2 
\sqrt{\sqrt{p}\left (pN-\frac{1}{2} \right )} 
\sqrt{\frac{u}{\sigma_{U}}} \right )\nonumber\\
\label{eq:pdfU_BesselK}
\label{eq:samplepdfU_final}
\label{eq:samplepdfU_incoh_BesselK}
\eea
with normalization constant $C_U$ given by
\bea
C_U 
&\stackrel{\Delta}{=}&
\frac{2}
     {\Gamma(p) \Gamma\left (pN-\frac{1}{2}\right )}
     p^{\frac{1}{4} \left [ p(N+1)-\frac{1}{2} \right ]} \nonumber\\
     &~&\times \left ( pN-\frac{1}{2} \right )^{\frac{1}{2} \left [ p(N+1)-\frac{1}{2} \right ]}.
     \label{eq:CpU}
\eea
Again, (\ref{eq:samplepdfU_incoh_BesselK}) is one of McKay's Bessel $K$ distributions.
It is remarkable that for the case $p=1$, the same type of Bessel $K$ distribution (\ref{eq:samplepdfU_incoh_BesselK}) has been obtained in the context of sea echo for microwave radar \cite{jake1}, but as an {\em ensemble\/} pdf of $U_\alpha$ based on different starting assumptions, viz., for a finite random walk in the complex plane that assumes a randomly fluctuating and large but finite number of steps (independent scattering contributions) that is distributed according to a negative binomial distribution, as a discretization of a gamma distribution for continuous stepping. 
For general values of $p$, the pdf (\ref{eq:samplepdfU_incoh_BesselK}) was also obtained in \cite{arnaKdf} as a limit distribution for imperfect reverberation based on a Bayesian model for a physical process of omnidirectional scattering. 
Hence, the present derivation of (\ref{eq:samplepdfU_incoh_BesselK}) shows that Bessel $K$ distributions are far more universal than previously thought, as they can arise under the much less restrictive condition of a mere small-sample effect for an underlying Gaussian field, as opposed to a need for any functional form of {\it a priori} distributions for a fluctuating $\nu$, whether chosen {\it ad hoc} or otherwise. 
However, any apparent departure from an underlying Gauss normal field distribution must be interpreted with due care and does not necessarily point to physical nonlinearity. The identification of Bessel $K$ distributions as sampling distributions for small sample sets of received power is supported by measurements of the evolution of the distribution function of pulsed energy in a reverberant cavity \cite{arnaPRL}. Such evolution is characterized by a steady growth in the number of multipath components and, hence, $\nu$.

Figs. \ref{fig:pdfP_paramNt_closedform_homog_incoh_Cart}a,
\ref{fig:pdfP_paramNt_closedform_homog_incoh_plan}a, and \ref{fig:pdfP_paramNt_closedform_homog_incoh_tot}a show the pdf (\ref{eq:pdfU_BesselK}) of incoherently detected Cartesian ($p=1$), planar ($p=2$) and total ($p=3$) field intensities, respectively, for selected values of $N$. The heavier tail and the sharper peak (mode) of the distribution at smaller values of $N$ are characteristic features.
Fig. \ref{fig:sigmaP}a shows the standard deviation of (\ref{eq:pdfU_BesselK}) for these three cases.

\subsubsection{Standardized intensity\label{sec:power_incoh_stand}}
For comparison with the empirical pdf, we consider the sampling pdf of the ratio of the two random variables $U$ and $S_U$, viz.,
\bea
W\stackrel{\Delta}{=}\frac{U}{S_U}.
\eea
The calculation of $f_{W}(w;N)$ is detailed in Sec. \ref{app:samplepdfW}. The result is a Fisher-Snedecor $F$ distribution
\bea
f_{W}(w;N) 
&=& \frac{C_W}{s_{W}}
\left ( \frac{w}{s_{W}}\right )^{p-1} \nonumber\\
&~&\times \left ( 1 + \frac{\sqrt{p}}{pN-\frac{1}{2}} \thinspace w \right )^{-[p(N+1)-\frac{1}{2}]}
\label{eq:samplepdfW_incoh_FisherF}
\eea
as a counterpart of (\ref{eq:samplepdfX_coh_StudentT}), where
\bea
C_{W} \stackrel{\Delta}{=}
\frac{p^{p/2} }{\left ( pN-\frac{1}{2}\right )^{p}}
\frac{\Gamma\left ( pN-\frac{1}{2}+p \right )}{\Gamma(pN-\frac{1}{2})\Gamma(p)}.
\label{eq:defCW}
\eea

\begin{figure}[htb] \begin{center} \begin{tabular}{cc} \ 
\epsfxsize=7cm \epsfbox{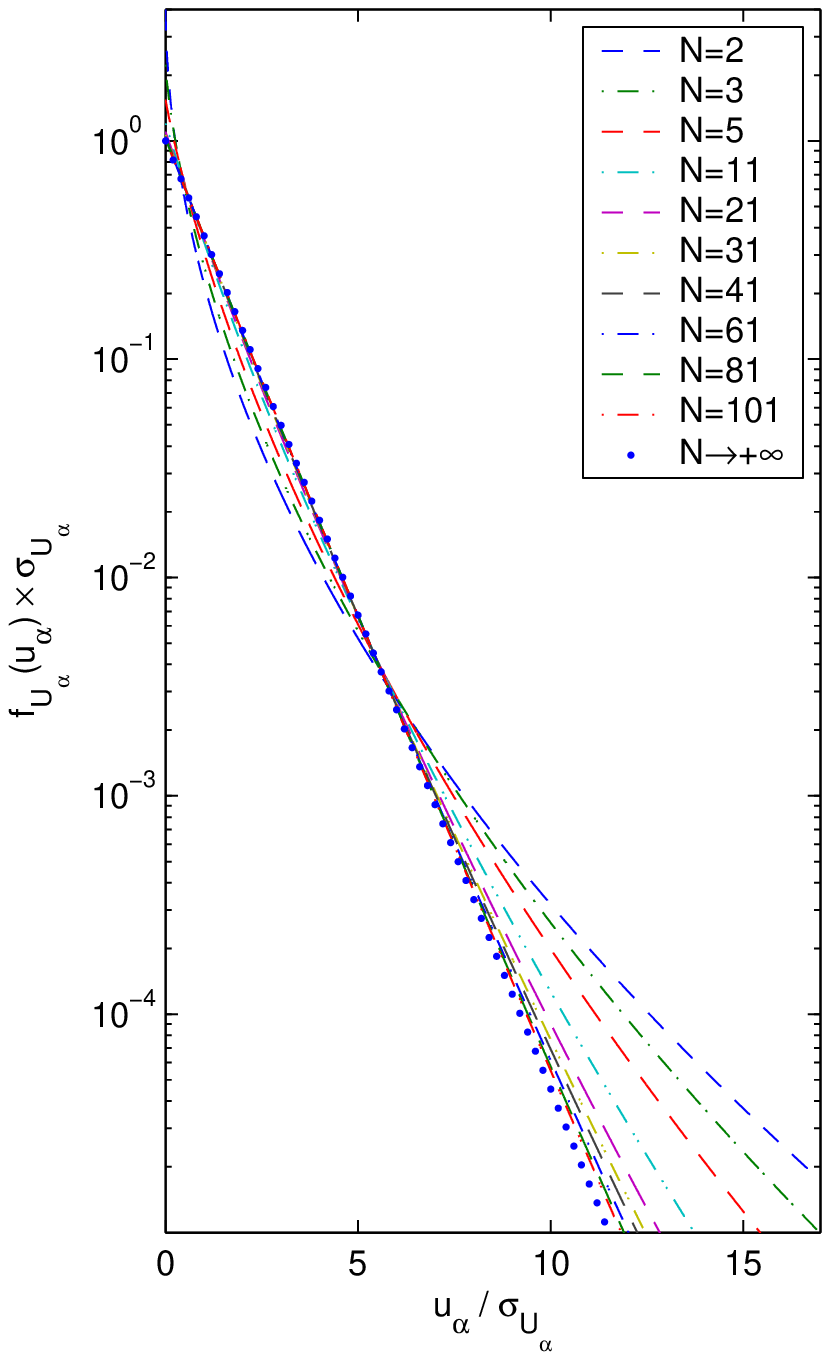}\ & \epsfxsize=7cm \epsfbox{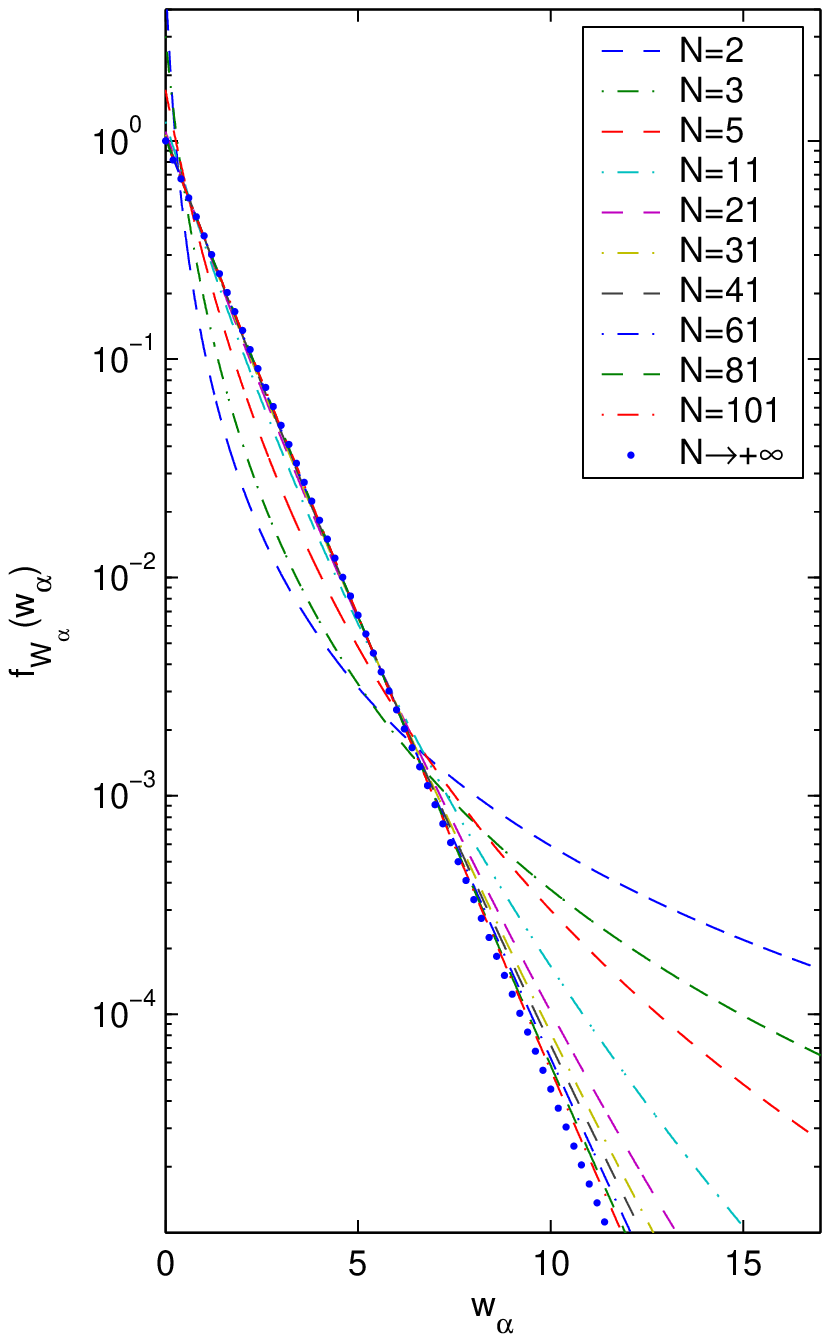}\ \\
\end{tabular} \end{center}
\caption{\label{fig:pdfP_paramNt_closedform_homog_incoh_Cart}
\small
(color on-line)
Sampling probability density function for intensity or energy density of Cartesian component of field ($p{=}1$) at selected values of $N$ based on incoherent detection:
(a) Bessel $K$ sampling pdfs of $U_\alpha$
[eq. (\ref{eq:samplepdfU_incoh_BesselK})]; 
(b) Fisher-Snedecor $F$ sampling pdfs  of $W_\alpha=U_\alpha/S_{U_\alpha}$
[eq. (\ref{eq:samplepdfW_incoh_FisherF})].
In both plots, the right tail becomes thinner for increasing values of $N$.
}
\end{figure}

\begin{figure}[htb] \begin{center} \begin{tabular}{cc} \ 
\epsfxsize=7cm \epsfbox{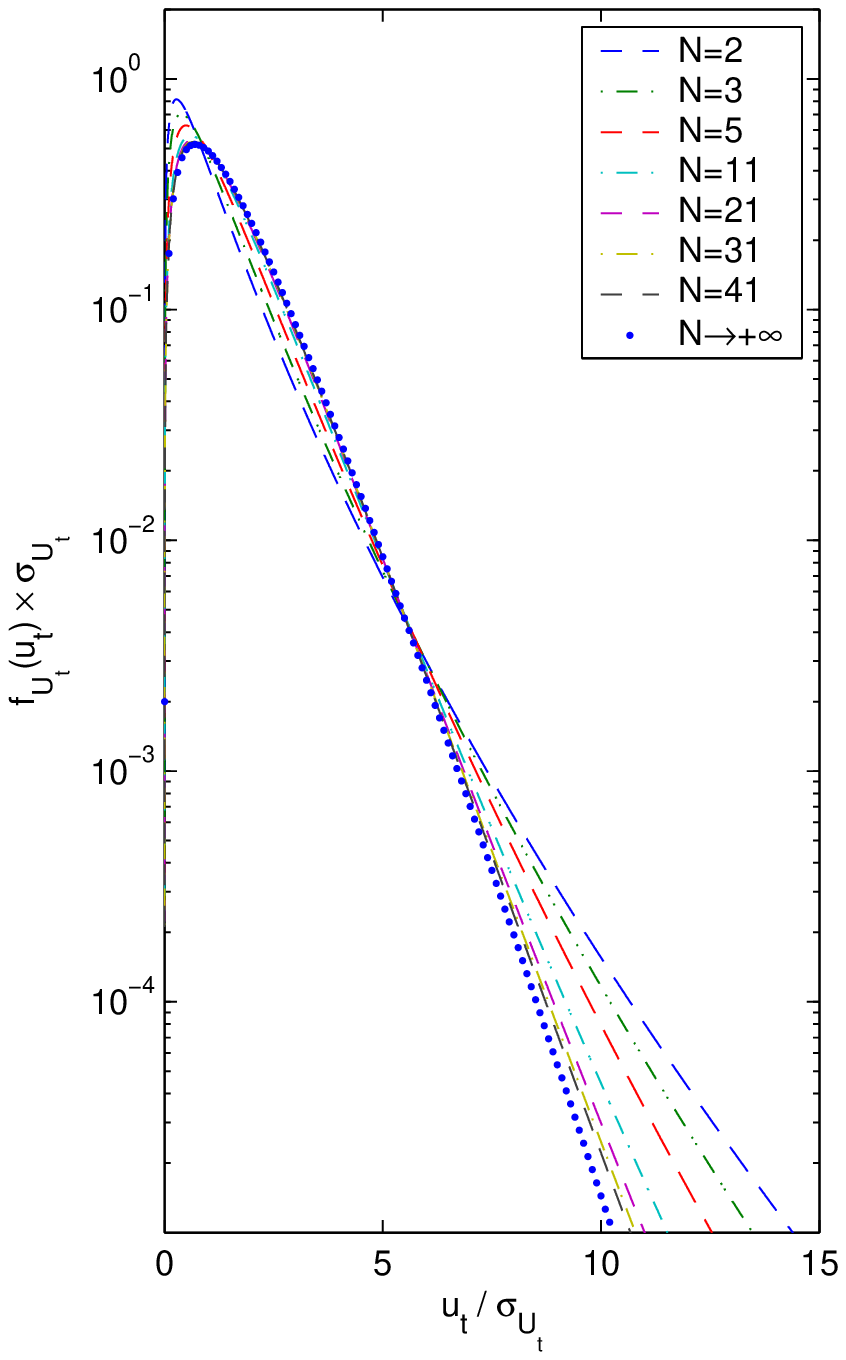}\ & \epsfxsize=7cm \epsfbox{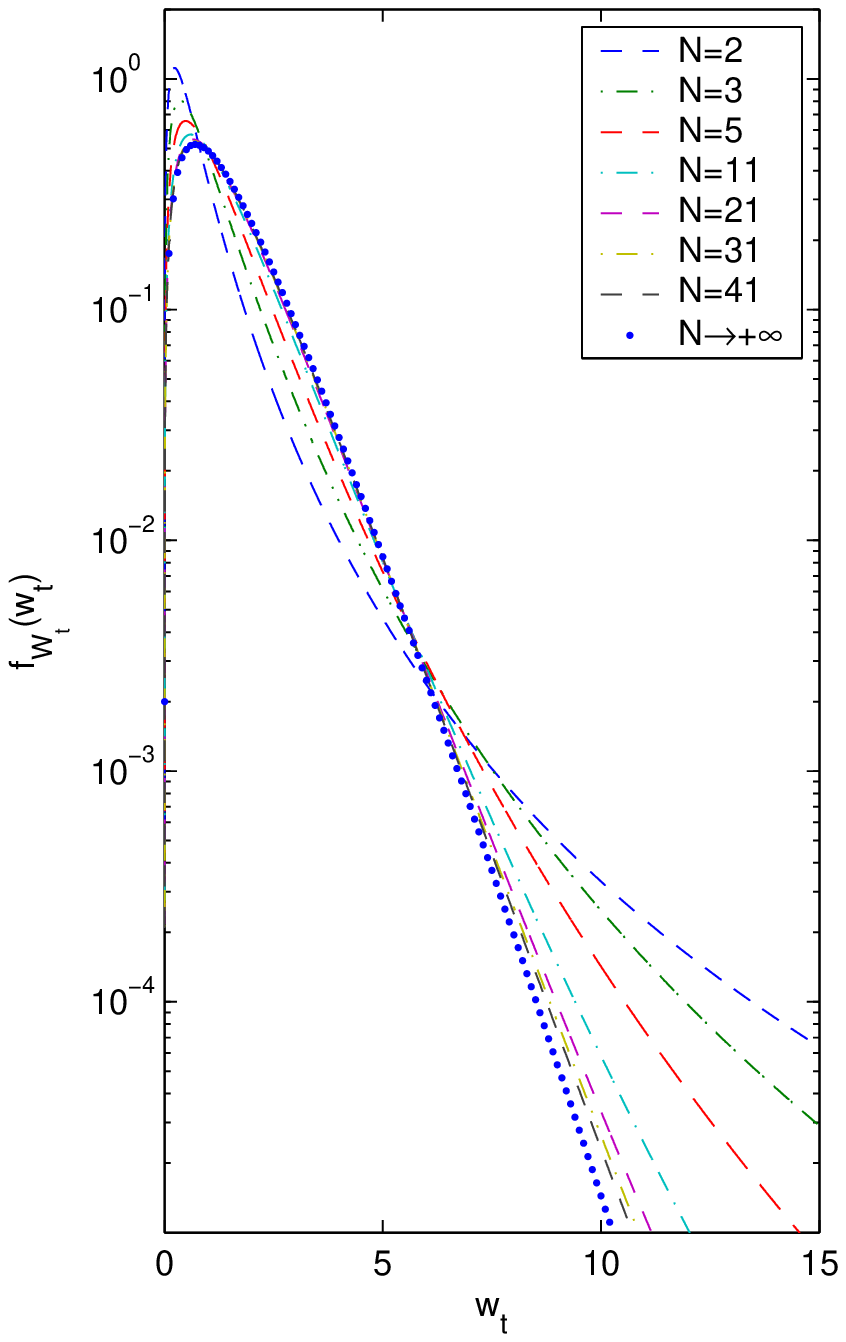}\ \\
\end{tabular} \end{center}
\caption{\label{fig:pdfP_paramNt_closedform_homog_incoh_plan}
\small
(color on-line)
Probability density function of intensity or energy density of planar field ($p{=}2$) at selected values of $N$ based on incoherent detection:
(a) Bessel $K$ sampling pdfs  of $U_{\rm t}$
[eq. (\ref{eq:samplepdfU_incoh_BesselK})]; 
(b) Fisher-Snedecor $F$ sampling pdfs  of $W_{\rm t}=U_{\rm t}/S_{U_{\rm t}}$
[eq. (\ref{eq:samplepdfW_incoh_FisherF})].
In both plots, the right tail becomes thinner for increasing values of $N$.
}
\end{figure}

\begin{figure}[htb] \begin{center} \begin{tabular}{cc} \ 
\epsfxsize=7cm \epsfbox{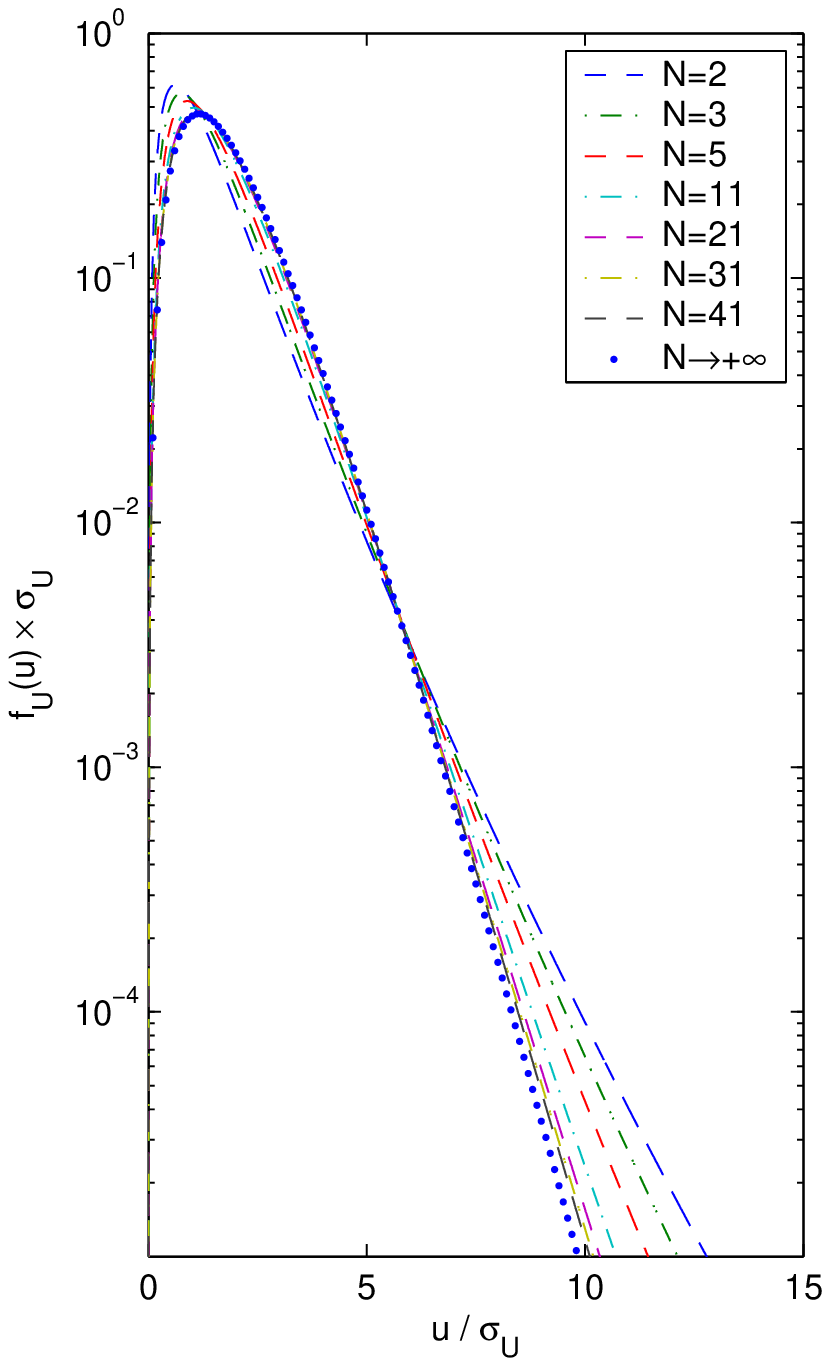}\ & \epsfxsize=7cm \epsfbox{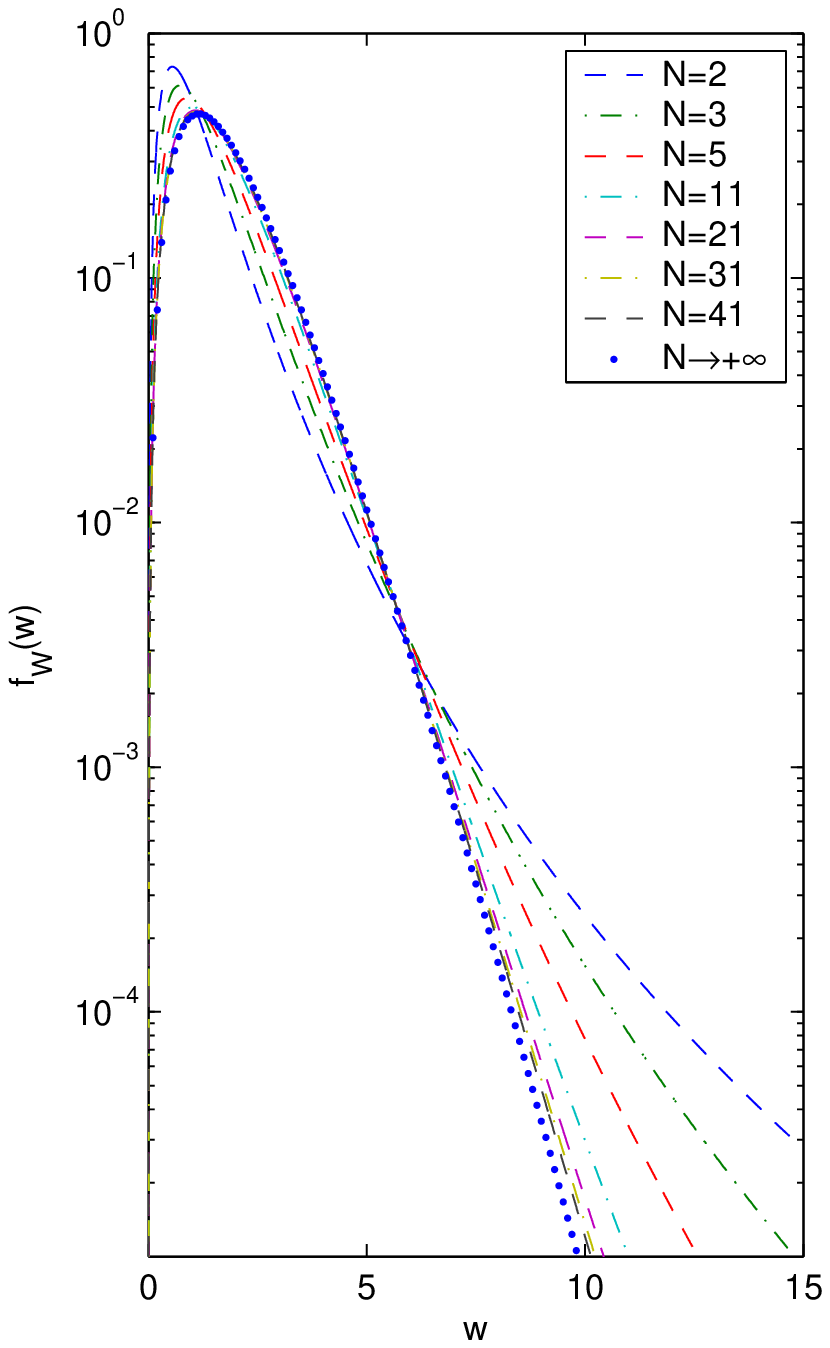}\\\
\end{tabular} \end{center}
\caption{\label{fig:pdfP_paramNt_closedform_homog_incoh_tot}
\small
(color on-line)
Probability density function of intensity or energy density of total (vector) field ($p{=}3$) at selected values of $N$ based on incoherent detection:
(a) Bessel $K$ sampling pdfs  of $U$ [eq. (\ref{eq:samplepdfU_incoh_BesselK})]; 
(b) Fisher-Snedecor $F$ sampling pdfs  of $W=U/S_U$ [eq. (\ref{eq:samplepdfW_incoh_FisherF})].
In both plots, the right tail becomes thinner for increasing values of $N$.
}
\end{figure}

\begin{figure}[htb] \begin{center} \begin{tabular}{cc} \ 
\epsfxsize=7.5cm \epsfbox{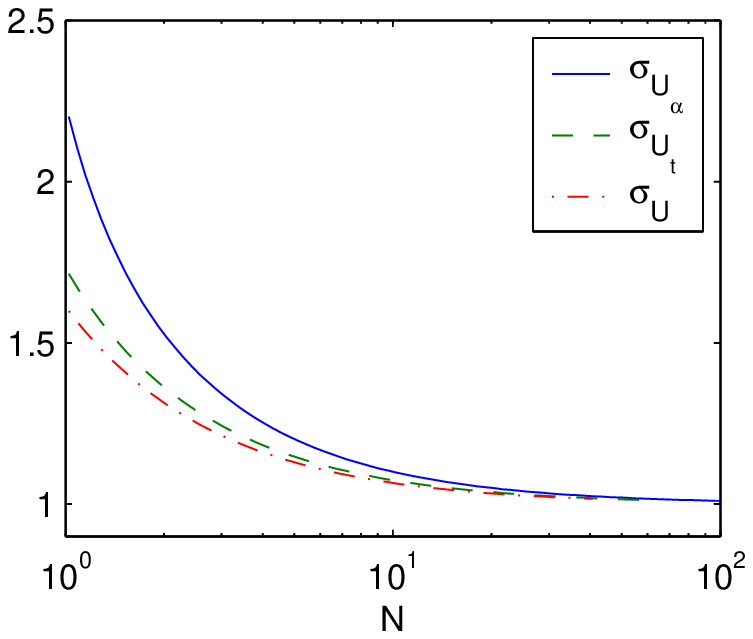}\ &
\epsfxsize=7.5cm \epsfbox{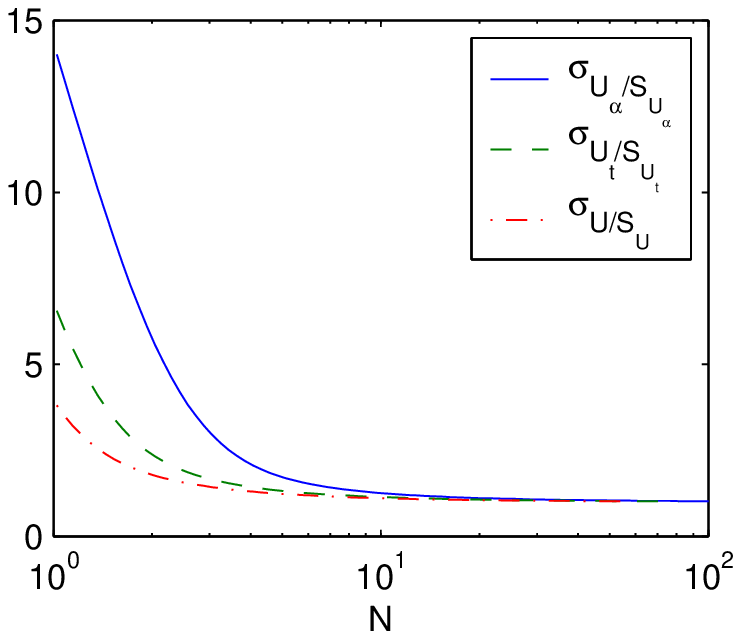}\ \\
(a) & (b)
\end{tabular} \end{center}
\caption{\label{fig:sigmaP}
\small
(color on-line)
Sampling standard deviations as a function of $N$
(a) for Bessel $K$ sampling distributions (\ref{eq:samplepdfU_incoh_BesselK}) of $U_\alpha$, $U_{\rm t}$, and $U$;
(b) for Fisher-Snedecor $F$ sampling distributions (\ref{eq:samplepdfW_incoh_FisherF}) of $U_\alpha/S_{U_\alpha}$, and $U_{\rm t}/S_{U_{\rm t}}$, $U/S_U$. }
\end{figure}

The pdf (\ref{eq:samplepdfW_incoh_FisherF}) for the incoherently detected and sample-standardized Cartesian ($p=1$), planar ($p=2$) and total ($p=3$) field intensities $W_\alpha$, $W_{\rm t}$, and $W$ is shown in Figs. \ref{fig:pdfP_paramNt_closedform_homog_incoh_Cart}b, \ref{fig:pdfP_paramNt_closedform_homog_incoh_plan}b, and \ref{fig:pdfP_paramNt_closedform_homog_incoh_tot}b, respectively, at selected values of $N$.
Comparison between the $K$ distributed intensities with their $F$ distributed sample-standarized values in Figs. \ref{fig:pdfP_paramNt_closedform_homog_incoh_Cart}--\ref{fig:pdfP_paramNt_closedform_homog_incoh_tot} shows that, for a given value of $N$, the sampling pdf of $W$ exhibits a more pronounced spread than for $U$, owing to the larger uncertainty of $W$ caused by fluctuations of $S_{U}$. 
Fig. \ref{fig:sigmaP}b shows the standard deviation of (\ref{eq:samplepdfW_incoh_FisherF}) for the three cases, indicating much larger standard deviations for small $N$ compared to those in Fig. \ref{fig:sigmaP}a for nonstandardized intensities.

\subsubsection{Non-standardized vs. non-normalized intensity\label{sec:power_incoh_nonnorm}}
In Sec. \ref{sec:power_incoh_nonstand}, $f_U(u;N)$ was obtained via the sample-{\em standardized\/} variate $U/S_U$ in the cpdf (\ref{eq:cpdf_incoh_approx}).
However, since $\mu_U \not = 0$, we could have equally used the sample-{\em normalized\/} variate $U/M_U$ for the cpdf to reference the data in this case. 
To this end, instead of (\ref{eq:pdfU_nonstand_gen}), we now use
\bea
f_U(u;N) &=&
\int^{+\infty}_0 f_{U|M_U}(u|m_U) f_{M_U}(m_u;N) {\rm d}m_U~~~~
\label{eq:pdfU_nonsnorm_gen}
\eea
with \cite{arnaTQE2}
\bea
f_{U|M_U}(u|m_U) &=& \frac{p^{p}}{\Gamma(p) ~ m_{U}} \left ( \frac{u}{m_U} \right )^{p-1} \exp \left ( - {p} \frac{u}{m_{U}} \right )~~~~~
\label{eq:cpdf_incoh_approx_nonnorm}\\
f_{M_U}(m_U; N)
&=&
\frac{\left( pN \right)^{\frac{pN}{2}-1}}
{\Gamma\left ( pN \right )\thinspace \sigma_{M_U}}
\left ( \frac{m_U}{\sigma_{M_U}} \right )^{pN-1} \nonumber\\
&~&\times \exp \left ( - \sqrt{pN}~\frac{ m_U}{\sigma_{M_U}} \, \right )
\label{eq:pdf_mP_exact}
\eea
and $\mu_U = m_U$,
$\sigma_U = \sqrt{N} \sigma_{M_U}
$,
whence (\ref{eq:pdfU_nonsnorm_gen}) becomes
\bea
f_U(u;N)
&=&
\frac{C^\prime_U}{\sigma_{U}} \left (  \frac{u}{\sigma_{U}} \right )^{\frac{1}{2}p(N+1) - 1} 
\nonumber\\ 
&~& \times K_{p(N-1)} \left ( 2 
p^{\frac{3}{4}} \sqrt{N} 
\sqrt{\frac{u}{\sigma_{U}}} \right )~~~~
\label{eq:pdfU_BesselK_nonnorm}
\label{eq:samplepdfU_final_nonnorm}
\label{eq:samplepdfU_incoh_BesselK_nonnorm}
\eea
with normalization constant $C^\prime_U$ given by
\bea
C^\prime_U \stackrel{\Delta}{=}
\frac{2~p^{\frac{3}{4} p(N+1)} ~N^{\frac{1}{2} p(N+1)}}
     {\Gamma(p) \Gamma\left (pN\right )}   .
     \label{eq:CppU}
\eea
The form (\ref{eq:pdfU_BesselK_nonnorm}) is readily re-expressed in terms of $\mu_U$ by replacing $\sigma_U$ with $\mu_U/\sqrt{p}$.
Comparing (\ref{eq:pdfU_BesselK_nonnorm})--(\ref{eq:CppU}) to (\ref{eq:pdfU_BesselK})--(\ref{eq:CpU}) shows that the non-normalized cpdf (\ref{eq:pdfU_BesselK_nonnorm}) is retrieved by replacing $N-(2p)^{-1}$ in the non-standardized cpdf (\ref{eq:pdfU_BesselK}) by $N$, i.e., resulting in a marginal increase of the number of independent samples. In other words, the non-normalized cpdf is marginally closer to the asymptotic ensemble pdf ($N\rightarrow+\infty$) than the non-standardized cpdf, resulting in slightly smaller uncertainties. This result is made plausible by the fact that the uncertainty of the sampling mean value is smaller than for the sampling standard deviation. It is worth emphasizing that, even though $f_U(u;N)$ is for the (dimensioned) energy density $U$, the chosen route for arriving at this pdf -- i.e., whether via intermediary standardization or normalization in the cpdf of $U$ -- has an effect on the number of degrees of freedom of the end result but not on the functional form.
In summary, whenever $pN \gg 1$ (i.e., for most practical cases), standardization and normalization yield indistinguishable final results.

\subsubsection{Intensity of biased field}
Instead of (\ref{eq:cpdf_incoh_approx}), the sampling cpdf of $U|S_{U}$ for the incoherently detected intensity of a biased field is a generalization of the so-called modified Nakagami--Rice $m$ distribution, given in self-sufficient form by
\bea
f_{U_\alpha|S_{U_\alpha}}(u_\alpha|s_{U_\alpha}) &=& \frac{\sqrt{1+2k_{U_\alpha}}}{s_{U_\alpha}} \nonumber\\
&~&\times \exp \left ( - \sqrt{1+2k_{U_\alpha}} \frac{u_\alpha+u_{\alpha 0}}{s_{U_\alpha}} \right ) \nonumber\\
&~&\times I_0\left ( 2 \sqrt{1+2k_{U_\alpha}} \frac{\sqrt{u_{\alpha 0}u_\alpha}}{s_{U_\alpha}} \right )
\label{eq:cpdf_incoh_bias}
\eea
with
\bea
k_{U_\alpha}\stackrel{\Delta}{=}\frac{|E_{\alpha 0}|^2}{\langle |E_\alpha-E_{\alpha 0}|^2\rangle}= \frac{1-n^2_{U_\alpha} + \sqrt{1-n^2_{U_\alpha}}}{n^2_{U_\alpha}}
\eea
where $n_{U_\alpha}$ is a sample value of $\nu_{U_\alpha} \stackrel{\Delta}{=}\sigma_{U_\alpha}/\mu_{U_\alpha}$.
Since $\sigma_{U_\alpha}=\sigma_{U_\alpha-u_{\alpha 0}}$ for any constant (deterministic) value $u_{\alpha 0}$, the pdf $f_{S_{U_\alpha}}(s_{U_\alpha}; N)$ is still given by (\ref{eq:pdf_stdP_exact}) with $p=1$. Thus,
\bea
f_{U_\alpha}(u_\alpha;N) 
&=& \frac{C^\prime_{U_\alpha}}{\sigma_{U_\alpha}}
\int^{+\infty}_0 
\sqrt{1+2k_{U_\alpha}}
\left ( \frac{s_{U_\alpha}}{\sigma_{{U_\alpha}}} \right )^{N-\frac{5}{2}} 
\nonumber\\ &~&\times 
\exp \left [ - \left ({N - \frac{1}{2}}\right )~\frac{ s_{U_\alpha}}{\sigma_{{U_\alpha}}} \, \right ]\nonumber\\
&~&\times
\exp \left ( - \sqrt{1+2k_{U_\alpha}} \frac{u_\alpha+u_{\alpha 0}}{s_{U_\alpha}} \right ) \nonumber\\
&~&\times I_0\left ( 2 \sqrt{1+2k_{U_\alpha}} \frac{\sqrt{u_{\alpha 0}u_\alpha}}{s_{U_\alpha}} \right )
{\rm d}s_{U_\alpha}.~~~~~
\label{eq:biasU}
\eea
Note that $k_{U_\alpha}$ depends implicitly on $s_{U_\alpha}$: to first approximation, $k_{U_\alpha} \simeq s_{U_\alpha} / \mu_{U_\alpha}$.

\subsection{Coherent detection}
\begin{figure}[htb] \begin{center} \begin{tabular}{c} \ 
\epsfxsize=7cm \epsfbox{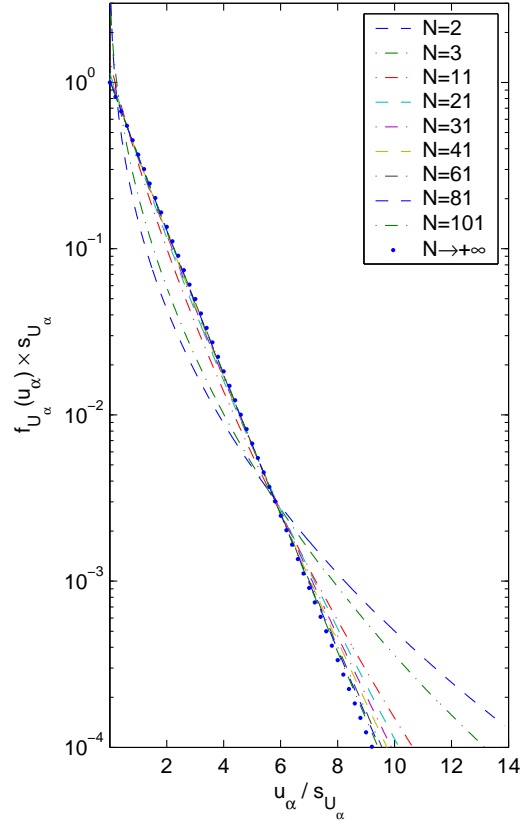}\ 
\end{tabular} \end{center}
\caption{\label{fig:pdfP_paramNt_coh_coh_Cart_BesselK_log}
\small
(color on-line)
Probability density function of intensity or energy density of Cartesian component of field ($p{=}1$) derived from coherent detection at selected values of $N$:
Bessel $K$ sampling pdf of $U_\alpha$ [eqn. (\ref{eq:samplepdfU_coh_BesselK})].
}
\end{figure}

In the case of estimating $f_U(u;N)$ from measurements of $E$, the sampling pdf takes a different form from that for incoherent detection, as we show next.
Knowing the pdf of $E_\alpha=E^\prime_\alpha-\rmj E^{\prime\prime}_\alpha$, we can derive the pdf of $U = |E|^2 = p |E_\alpha|^2$, where $|E_\alpha|^2\equiv {E^\prime_\alpha}^2 + {E^{\prime\prime}_\alpha}^2 = U_\alpha$ is the intensity of a Cartesian field component, and similarly for the electric or magnetic power $P \propto U$.
Since $f_{U^{\prime(\prime)}_\alpha}(u^{\prime(\prime)}_\alpha) \propto f_{E^{\prime(\prime)}_\alpha}(e^{\prime(\prime)}_\alpha = \sqrt{u^{\prime(\prime)}_\alpha}) / \sqrt{u^{\prime(\prime)}_\alpha}$, we obtain
\bea
f_{U_\alpha}(u_\alpha;N) 
&=& C_{U_\alpha} \int^{u_\alpha}_0 f_{U^\prime_\alpha}(x;N) f_{U^{\prime\prime}_\alpha}(u_\alpha-x;N) {\rm d}x\label{eq:fUtemp}\\
&=& C_{U_\alpha} \int^{u_\alpha}_0 \frac{f_{E^\prime_\alpha}(\sqrt{x};N) f_{E^{\prime\prime}_\alpha} (\sqrt{u_\alpha-x};N) }{ \sqrt{x}\sqrt{u_\alpha-x}} {\rm d}x\nonumber\\
\label{eq:fUCart}
\label{eq:samplepdfU_coh_BesselK}
\eea
where $C_{U_\alpha}$ is a normalization constant. 
The general expression (\ref{eq:fUCart}) makes allowance for the fact that $E^\prime_\alpha$ and $E^{\prime\prime}_\alpha$ may, in principle, have different pdfs (functionally and/or parametrically), although in most cases the pdf of $E_\alpha$ is circular, i.e., $f_{E^\prime_\alpha} = f_{E^{\prime\prime}_\alpha}$.
The sample pdfs of $U_{\rm t}\equiv U_x+U_y$ and $U\equiv U_x+U_y+U_z$ follow similarly from two- and threefold convolutions of (\ref{eq:fUCart}), respectively. Fig. \ref{fig:pdfP_paramNt_coh_coh_Cart_BesselK_log} shows the pdf (\ref{eq:fUCart}) for selected values of $N$. 

If ${U_\alpha}$ is to be compared with measured data, then the sampling pdf can be similarly calculated from $f_{X^\prime_\alpha}$ and $f_{X^{\prime\prime}_\alpha}$ as
\bea
f_{W_\alpha}(w_\alpha) 
&=& C_{W_\alpha} \int^{w_\alpha}_0 
\frac{f_{X^\prime_\alpha}(\sqrt{x};N) f_{X^{\prime\prime}_\alpha} (\sqrt{w_\alpha-x};N) }{ \sqrt{x}\sqrt{w_\alpha-x} } 
{\rm d}x \nonumber\\
\eea 
where $W_\alpha \stackrel{\Delta}{=} U_\alpha/S_{U_\alpha}$.

\section{Field amplitude\label{sec:amplitude}}
\subsection{Incoherent detection}
\subsubsection{Non-standardized field amplitude}
The sampling pdf of $A\stackrel{\Delta}{=} \sqrt{{E^\prime}^2+{E^{\prime\prime}}^2}$ follows in a manner similar to that for $U$. In this case, $A|S_A$ has a $\chi_{2p}$ cpdf, given in self-sufficient form as
\bea
f_{A|S_A}(a|s_A) &=& 
\frac{
2  
\left [ p - \left ( \frac{ \Gamma(p+\frac{1}{2}) }
                         { \Gamma(p) } 
            \right )^2 
\right ]^{p} \thinspace 
}
{\Gamma(p) \thinspace s_A}
\left ( \frac{a}{s_A} \right )^{2p-1} 
\nonumber\\&~&
\times
\exp \left \{ - \left [ p - \left ( \frac{ \Gamma(p+\frac{1}{2}) }
                                       { \Gamma(p) } 
                          \right )^2 
                \right ] 
                \left ( \frac{a}{s_A} \right )^2
     \right \}, \nonumber\\
\label{eq:field_one}
\eea
which has the Rayleigh distribution as a special case for $p=1$,
while $S_A$ has the $\chi_{2pN-1}$ distribution
\bea 
&~&\hspace{-8mm}
f_{S_A}(s_A; N)\nonumber\\
&=&
\frac{
2 \thinspace 
\left [ pN - \frac{1}{2} - \left ( \frac{ \Gamma(pN) }
                         { \Gamma(pN-\frac{1}{2}) } 
            \right )^2 
\right ]^{pN-\frac{1}{2}} 
}
{(pN -\frac{1}{2})~ \Gamma(pN-\frac{1}{2} ) \thinspace \sigma_{S_A} }
\left ( \frac{s_A}{\sigma_{S_A}} \right )^{2(pN-1)} 
\nonumber\\&~&
\times \exp \left \{ - \left [ pN -\frac{1}{2} - \left ( \frac{ \Gamma(pN) }
                                       { \Gamma(pN-\frac{1}{2}) } 
                          \right )^2 
                \right ]
                \left ( \frac{s_A}{\sigma_{S_A}} \right )^2
     \right \} \nonumber\\
\label{eq:pdf_stdE_exact}
\eea 
with $\sigma_{S_A}$ and $\sigma_A$ related via
\bea
\sigma_{S_A} = \sigma_A \sqrt{ 1 - \frac{1}{pN - \frac{1}{2}} \left ( \frac{\Gamma(pN)}{\Gamma \left ( pN - \frac{1}{2} \right )} \right )^2}.
\eea
The sampling pdf of $A$ is obtained as
a marginal pdf of $f_{A,S_A}(a,s_A;N)$ and is again a Bessel $K$ distribution, but of yet another type compared to (\ref{eq:samplepdfE_coh_BesselK})  and (\ref{eq:samplepdfU_incoh_BesselK}), viz.,
\bea
&~&\hspace{-6mm}
f_A(a;N) =
\int^{+\infty}_0 f_{A|S_A}(a|s_A) f_{S_A}(s_A;N) {\rm d}s_A\nonumber\\
&~& \hspace{-6mm} = \frac{C_A}{\sigma_{A}} \left ( \frac{a}{\sigma_{A}} \right )^{p(N+1)-\frac{3}{2}}\nonumber\\
&~& \hspace{-6mm} \times
K_{p(N-1)-\frac{1}{2}} 
\left ( 
2 \sqrt{ \left [ p  - \left ( \frac{\Gamma\left ( p + \frac{1}{2} \right )}{\Gamma (p)} \right )^2 \right ] 
         \left ( pN-\frac{1}{2} \right )} 
         \frac{a}{\sigma_{A}}
\right )
\nonumber\\
\label{eq:samplepdfA_incoh_BesselK}
\eea
where $C_A$ is obtained, with the aid of \cite[(6.561.16)]{grad1}, as
\bea
C_A &\stackrel{\Delta}{=}& 
\frac{4 }{\Gamma(p) \Gamma \left ( pN-\frac{1}{2} \right )}
\left [ p - \left ( \frac{\Gamma\left ( p + \frac{1}{2} \right )}{\Gamma (p)} \right )^2 \right ]^{\frac{1}{2} \left [ p(N +2)-1\right ]} \nonumber\\ &~& \times
\left ( pN - \frac{1}{2} \right )^{\frac{1}{2} \left [ p(N+1)-1 \right ]}.
     \label{eq:CA}
\eea
The pdf (\ref{eq:samplepdfA_incoh_BesselK}) for $p=1$ has been obtained in \cite{jake2} for the abovementioned scenario of a random walk with fluctuating number of steps with negative binomial distribution. For general $p$, this pdf has been retrieved as a limit distribution for imperfect reverberation in \cite{arnaKdf}.

\begin{figure}[htb] \begin{center} \begin{tabular}{cc} \ 
\epsfxsize=7cm \epsfbox{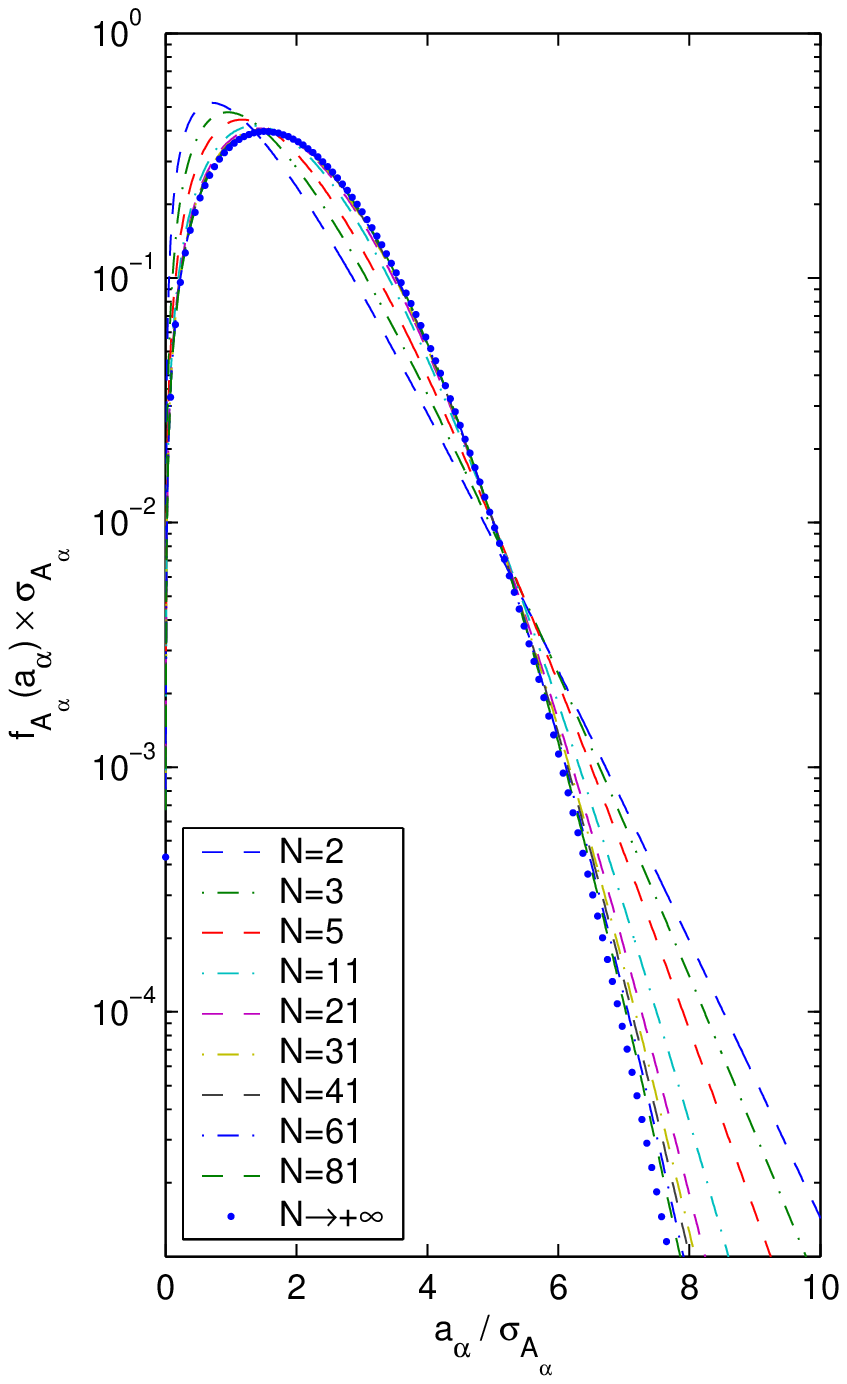}\ & \epsfxsize=7cm \epsfbox{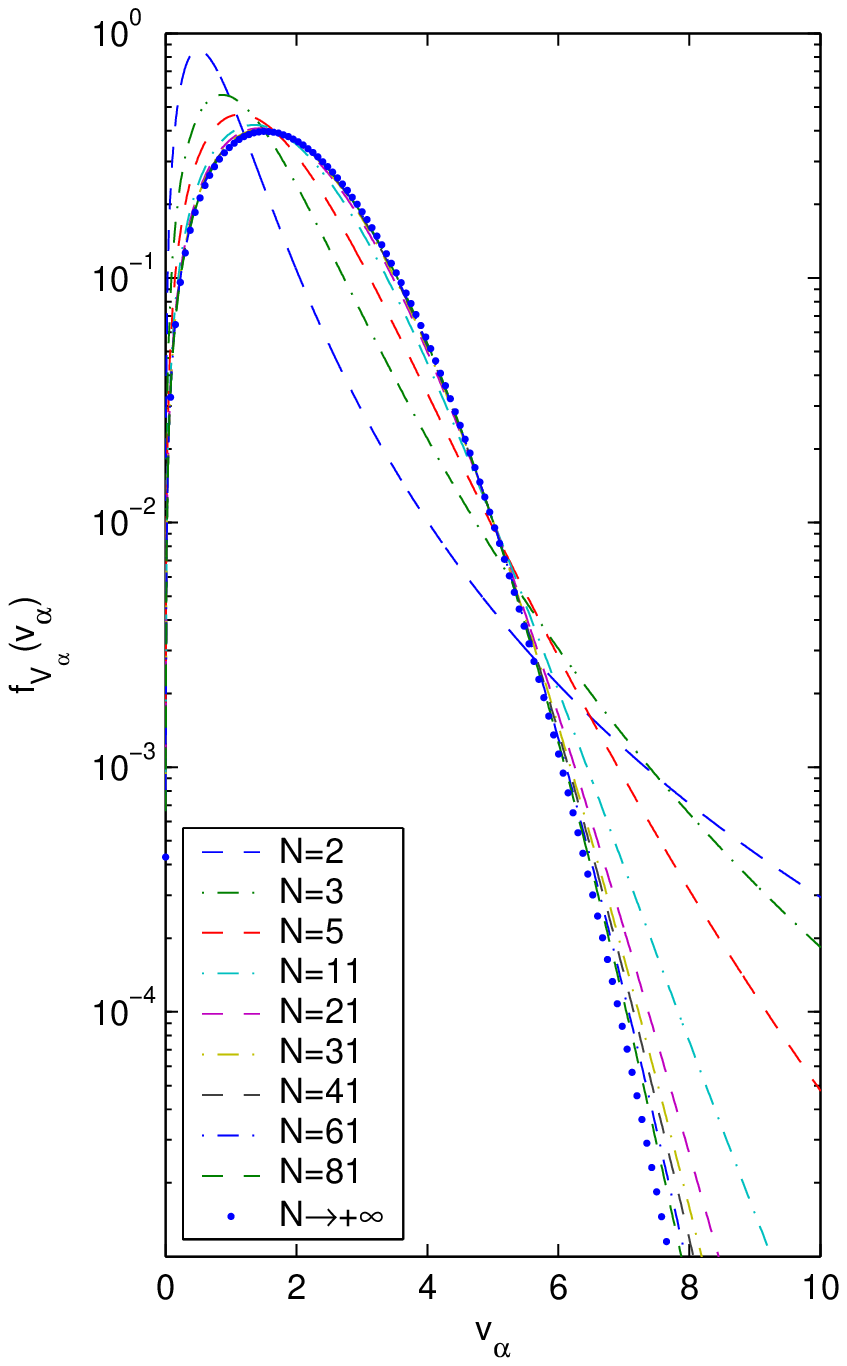}\ \\
\end{tabular} \end{center}
\caption{\label{fig:pdfA_paramNt_closedform_homog_incoh_Cart}
\small
(color on-line)
Sampling probability density function of amplitude of Cartesian component of field ($p=1$) at selected values of $N$ based on incoherent detection:
(a) Bessel $K$ sampling pdf (\ref{eq:samplepdfA_incoh_BesselK}) for $A_\alpha$; 
(b) root-$F$ pdf (\ref{eq:samplepdfV_final_copy}) for $V_\alpha=A_\alpha/S_{A_\alpha}$.
In both plots, the right tail becomes thinner for increasing values of $N$.
}
\end{figure}

\begin{figure}[htb] \begin{center} \begin{tabular}{cc} \ 
\epsfxsize=7cm \epsfbox{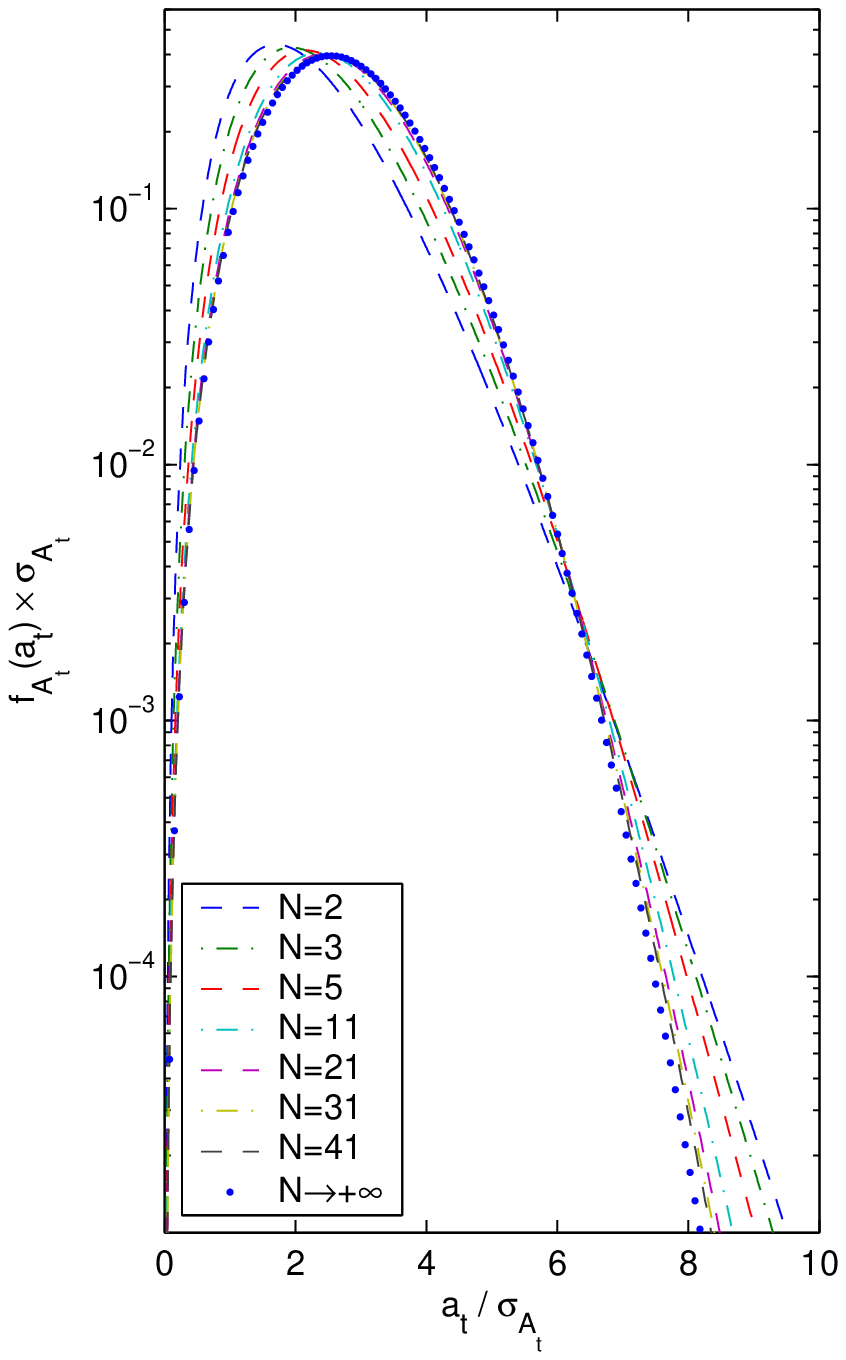}\ & \epsfxsize=7cm \epsfbox{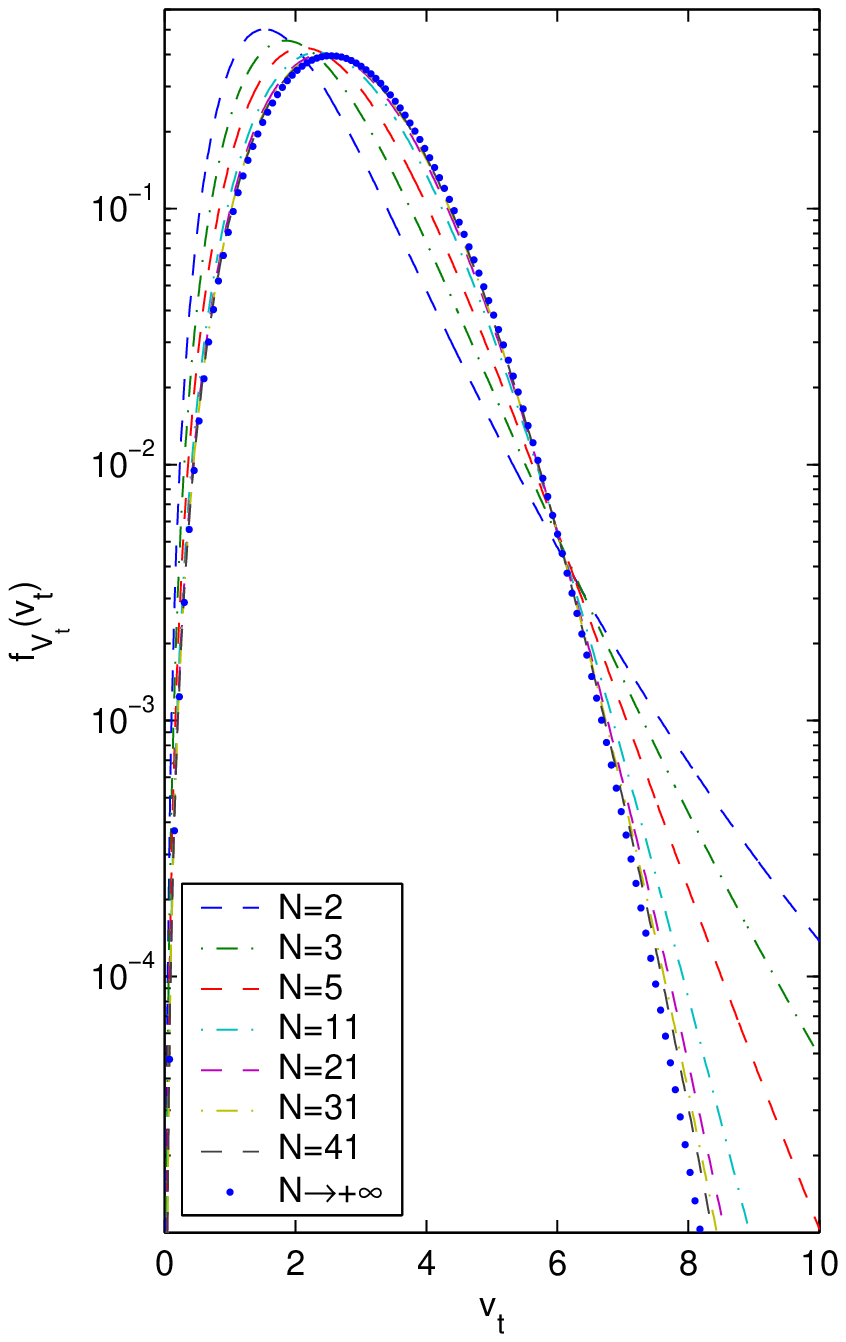}\ \\
\end{tabular} \end{center}
\caption{\label{fig:pdfA_paramNt_closedform_homog_incoh_plan}
\small
(color on-line)
Sampling probability density function of amplitude of planar field ($p{=}2$) at selected values of $N$ based on incoherent detection:
(a) Bessel $K$ pdf (\ref{eq:samplepdfA_incoh_BesselK}) for $A_{\rm t}$; 
(b) root-$F$ pdf (\ref{eq:samplepdfV_final_copy}) for $V_{\rm t}=A_{\rm t}/S_{A_{\rm t}}$.
In both plots, the right tail becomes thinner for increasing values of $N$.
}
\end{figure}

\begin{figure}[htb] \begin{center} \begin{tabular}{cc} \ 
\epsfxsize=7cm \epsfbox{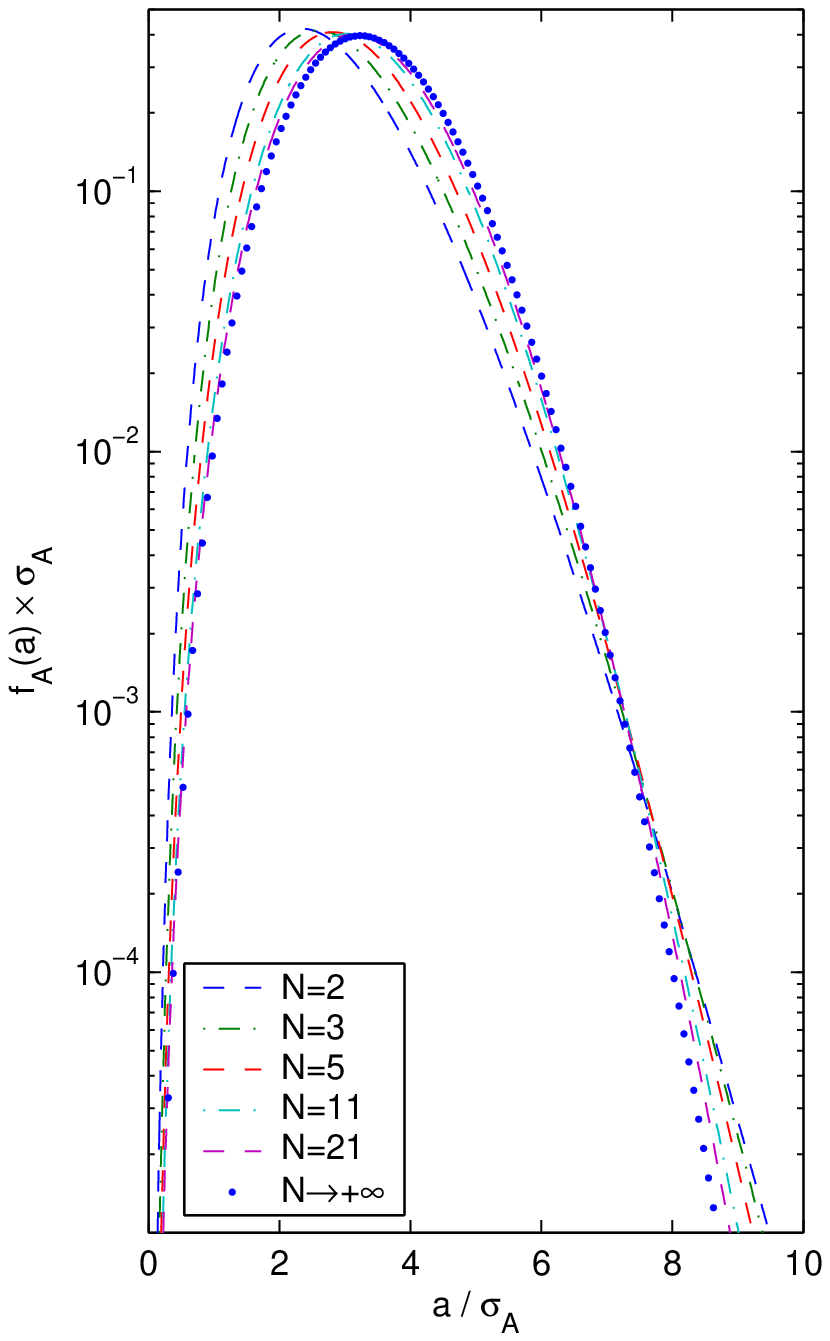}\ & \epsfxsize=7cm \epsfbox{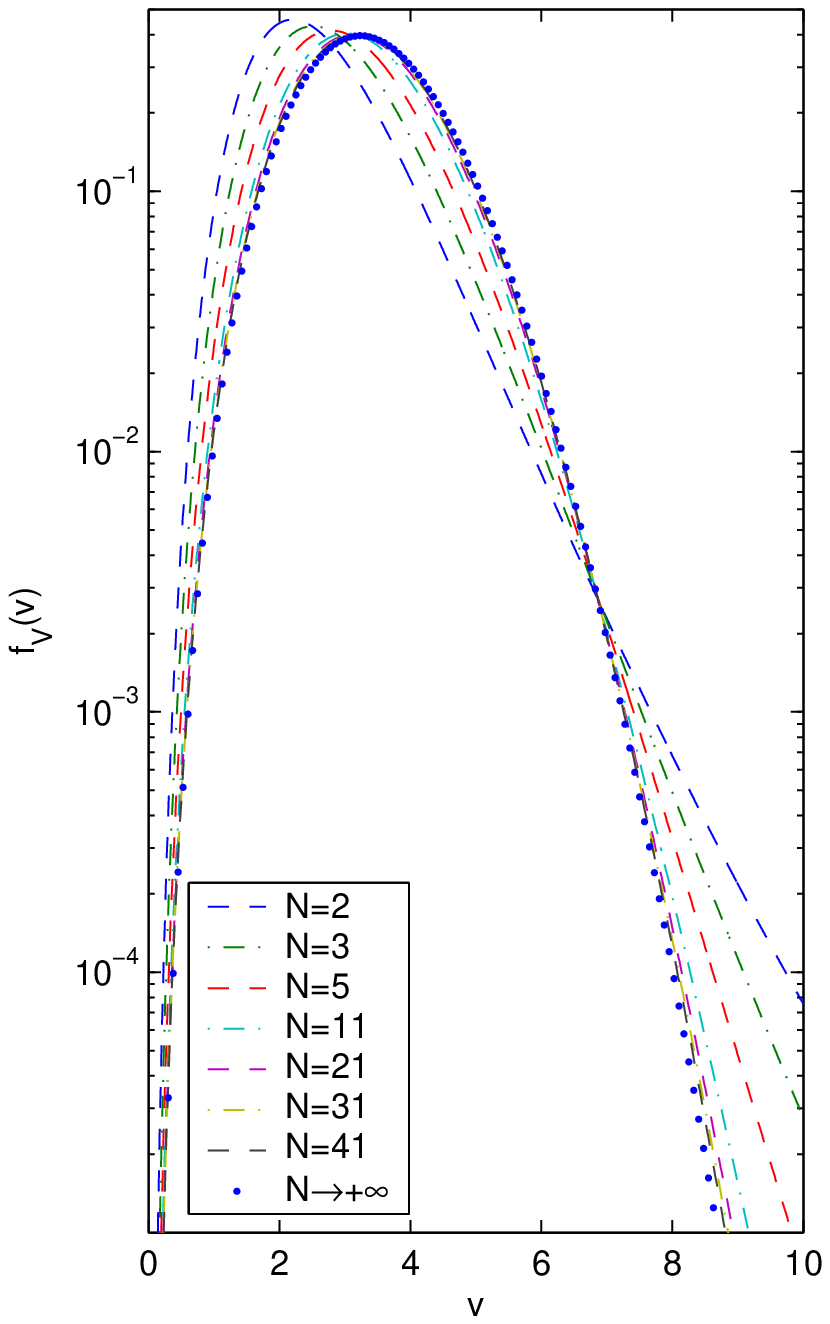}\\\
\end{tabular} \end{center}
\caption{\label{fig:pdfA_paramNt_closedform_homog_incoh_tot}
\small
(color on-line)
Sampling probability density function of amplitude of total (vector) field ($p{=}3$) at selected values of $N$ based on incoherent detection:
(a) Bessel $K$ pdf (\ref{eq:samplepdfA_incoh_BesselK}) for $A$; 
(b) root-$F$ pdf (\ref{eq:samplepdfV_final_copy}) for $V=A/S_{A}$.
In both plots, the right tail becomes thinner for increasing values of $N$.
}
\end{figure}

\begin{figure}[htb] \begin{center} \begin{tabular}{cc} \ 
\epsfxsize=7.5cm \epsfbox{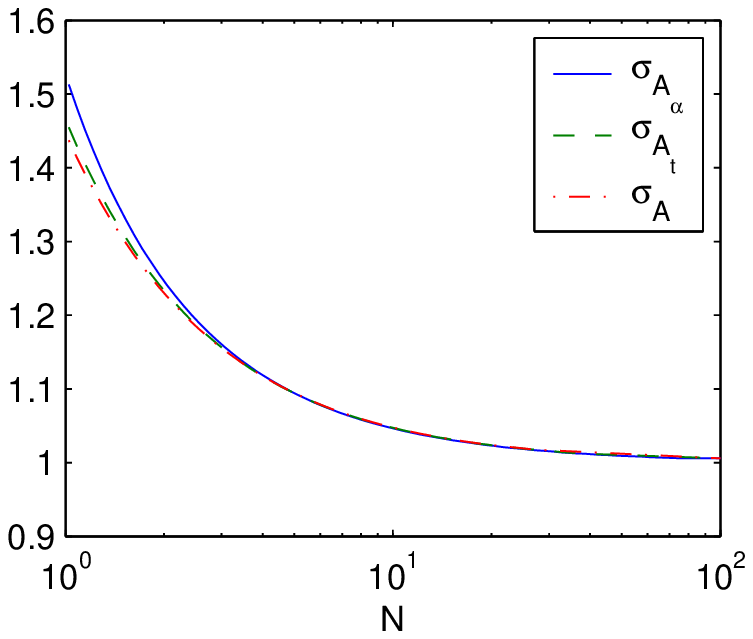}\ &
\epsfxsize=7.5cm \epsfbox{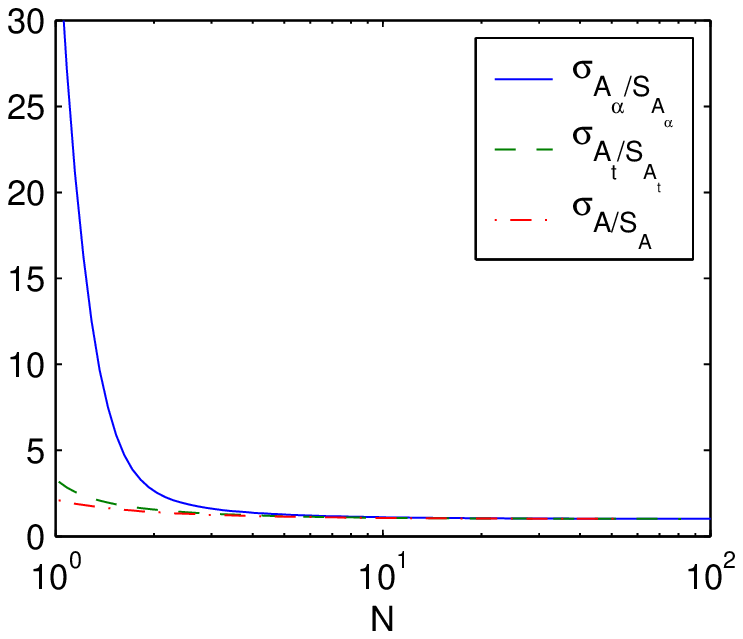}\ \\
(a) & (b)
\end{tabular} \end{center}
\caption{\label{fig:sigmaA}
\small
(color on-line)
Sampling standard deviation as a function of $N$
(a) for Bessel $K$ pdf (\ref{eq:samplepdfA_incoh_BesselK}) of $A_\alpha$, $A_{\rm t}$, and $A$;
(b) for Fisher-Snedecor $F$ pdf (\ref{eq:samplepdfV_final_copy}) of $A_\alpha/S_{A_\alpha}$, $A_{\rm t}/S_{A_{\rm t}}$, and $A/S_A$. }
\end{figure}

The pdf (\ref{eq:samplepdfA_incoh_BesselK}) for $p=1,2$ and $3$ is shown in Figs. 
\ref{fig:pdfA_paramNt_closedform_homog_incoh_Cart}a, 
\ref{fig:pdfA_paramNt_closedform_homog_incoh_plan}a, and 
\ref{fig:pdfA_paramNt_closedform_homog_incoh_tot}a, respectively, for selected values of $N$.
Fig. \ref{fig:sigmaA}a shows the corresponding standard deviations of (\ref{eq:samplepdfW_incoh_FisherF}).
Compared to Fig. \ref{fig:sigmaP}a, the increase of the standard deviations with decreasing $N$ is generally smaller and less dependent on dimensionality.

\subsubsection{Standarized field amplitude}
For comparison with the empirical pdf of a field amplitude measured by an electric or magnetic field probe, we consider the sampling pdf of the ratio of the variates $A$ and $S_A$, viz.,
\bea
V\stackrel{\Delta}{=}\frac{A}{S_A}.
\eea
The calculation of $f_V(v;N)$ is detailed in Sec. \ref{app:samplepdfV}, where the final result is shown to be
\bea
f_V(v;N) 
&=& C_V~
{v^{2p-1}} \nonumber\\
&~&\times {\left [ 1 + \frac{p-\left ( \frac{\Gamma\left ( p+\frac{1}{2} \right )}{\Gamma(p)}\right )^2}{pN-\frac{1}{2}} \thinspace v^2 \right ]^{-[p(N+1)-\frac{1}{2}]}}
\label{eq:samplepdfV_final_copy}
\eea
with
\bea
C_V \stackrel{\Delta}{=} \frac{2}{\left (pN-\frac{1}{2}\right )^{p}}
\left [ p - \left ( \frac{\Gamma\left ( p+\frac{1}{2} \right )}{\Gamma(p)}\right )^2 \right ]^{p} \frac{\Gamma\left (pN-\frac{1}{2}+p \right )}{\Gamma(pN-\frac{1}{2})\Gamma(p)}. \nonumber\\
\label{eq:defCV}
\eea
The sampling pdf (\ref{eq:samplepdfV_final_copy}) can be referred to as a root-$F$ distribution and
constitutes a counterpart of (\ref{eq:samplepdfX_coh_StudentT}) and (\ref{eq:samplepdfW_incoh_FisherF}).
Fig. \ref{fig:sigmaA}b shows its standard deviation.

\subsubsection{Amplitude of biased field}
Unfortunately, unlike (\ref{eq:cpdf_incoh_bias}), the cpdf $f_{A_\alpha|S_{A_\alpha}}(a_\alpha|s_{A_\alpha})$ cannot be expressed in self-sufficient closed form, because $\sigma_{A_\alpha}$ for the Nakagami--Rice cpdf of $A$ depends in a complicated manner on $\sigma_{E^{\prime(\prime)}}$. Therefore, we do not further pursue this case. A simpler approach is to derive $f_A(a;N)$ via variate transformation of (\ref{eq:biasU}).

\subsection{Coherent detection}
\begin{figure}[htb] \begin{center} \begin{tabular}{c} \ 
\epsfxsize=7cm \epsfbox{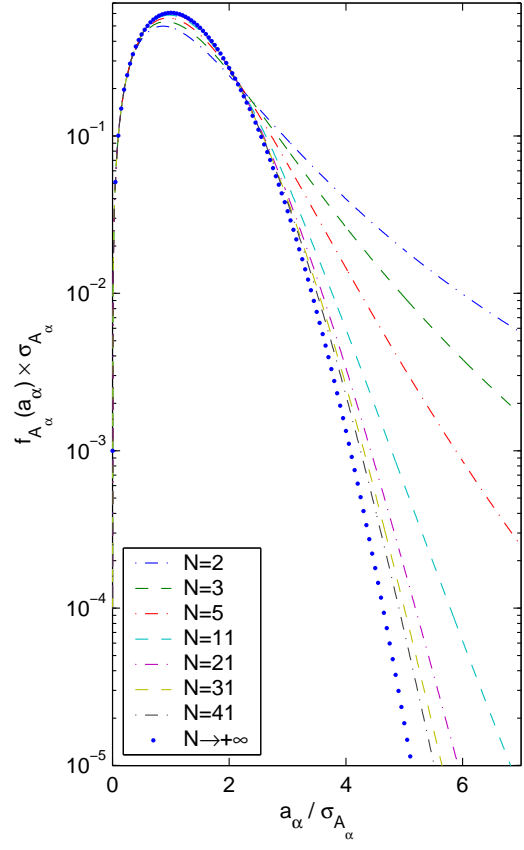}\ \\
\end{tabular} \end{center}
\caption{\label{fig:pdfA_paramNt_closedform_Cart}
\small
(color on-line)
Sampling probability density function of magnitude of Cartesian component of field ($p{=}1$) at selected values of $N$, derived for coherent detection. 
The right tail becomes thinner for increasing values of $N$.
}
\end{figure}

\begin{figure}[htb] \begin{center} \begin{tabular}{c} \ 
\epsfxsize=7cm \epsfbox{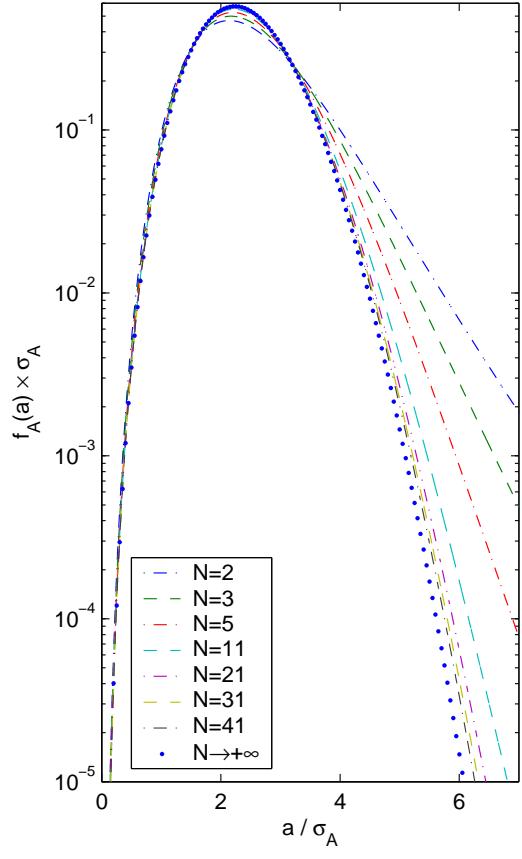}\ \\
\end{tabular} \end{center}
\caption{\label{fig:pdfA_paramNt_closedform_tot}
\small
(color on-line)
Sampling probability density function of magnitude of total (vector) field ($p{=}3$) at selected values of $N$, derived for coherent detection. 
In both plots, the right tail becomes thinner for increasing values of $N$.
}
\end{figure}

Using the variate transformation $A=\sqrt{U}$, the pdf of the local field magnitude $A = \sqrt{A^2_x+A^2_y+A^2_z}\equiv |E|\propto \sqrt{U}$ follows.
With $A_\alpha\stackrel{\Delta}{=}\sqrt{{E^\prime_\alpha}^2+{E^{\prime\prime}_\alpha}^2}$, (\ref{eq:fUCart}) yields
\bea
&~&\hspace{-8mm}
f_{A_\alpha}(a_\alpha;N) \nonumber\\
&=& a_\alpha \thinspace f_{U_\alpha}(u_\alpha=a^2_\alpha)\nonumber\\
&=& C_{A_\alpha} \thinspace a_\alpha \int^{a^2_\alpha}_0 \frac{f_{E^\prime_\alpha}(\sqrt{x};N) f_{E^{\prime\prime}_\alpha} (\sqrt{a^2_\alpha-x};N) }{ \sqrt{x}\sqrt{a^2_\alpha-x}} {\rm d}x ~~~~~
\label{eq:fA}
\eea
where $C_{A_\alpha}$ is a normalization constant. 
The pdf for $A_\alpha$ is shown for selected values of $N$ in Fig. \ref{fig:pdfA_paramNt_closedform_Cart} and for $A$ in Fig. \ref{fig:pdfA_paramNt_closedform_tot}.

In fact, one may consider the amplitude of the sampling field itself as a marginal of the joint sampling pdf $f_{E^\prime,E^{\prime\prime}}(e^\prime,e^{\prime\prime};N)$, followed by variate transformation to the amplitude $A=\sqrt{{E^\prime}^2+{E^{\prime\prime}}^2}$ and phase $\Phi=\tan^{-1}(E^{\prime\prime}/E^{\prime})$.
Assuming that $E^\prime$ and $E^{\prime\prime}$ are statistically independent (which, strictly, requires them to be normally distributed) so that $f_{E^\prime,E^{\prime\prime}}(e^\prime,e^{\prime\prime};N) = f_{E^\prime}(e^\prime;N)f_{E^{\prime\prime}}(e^{\prime\prime};N)$, then
\bea
f_A(a;N) &=& \frac{C_A}{\sigma_A} a^{{pN}-{1}} \int^\pi_{-\pi} \left ( {|\cos\phi-(a_0/a)|} {|\sin\phi|} \right )^{\frac{pN}{2}-1}\nonumber\\
&~&\times K_{\frac{pN}{2}-1} \left ( \sqrt{pN-1} \frac{|a\cos\phi-a_0|}{\sigma_{E^\prime}} \right )\nonumber\\
&~&\times
K_{\frac{pN}{2}-1} \left ( \sqrt{pN-1} \frac{|a\sin\phi|}{\sigma_{E^{\prime\prime}}} \right )
{\rm d}\phi
\eea
where $ a_0 \stackrel{\Delta}{=} \mu_{E^\prime} $ with $\mu_{E^{\prime\prime}}=0$.

\section{Extension to non-Gaussian ensemble distributions of the field: Iterative Bayesian scheme\label{sec:Bayes}}
From the theorem of total probability, it follows that
\bea
f_{S_{E^{\prime(\prime)}}|E^{\prime(\prime)}}
\propto
f_{E^{\prime(\prime)}|S_{E^{\prime(\prime)}}}
~
f_{S_{E^{\prime(\prime)}}}.
\label{eq:Bayes}
\eea
For non-Gaussian $E^{\prime(\prime)}$, for which the prior pdf $f_{S_{E^{\prime(\prime)}}}$ is more difficult to determine than in the exposition given before, (\ref{eq:Bayes}) can be used in an iterative process by means of an update equation for $f_{S_{E^{\prime(\prime)}}}$, where the latter can be assigned an initial $\chi_{N-1}$ distribution $f^{(0)}_{S_{E^{\prime(\prime)}}}$. This prior distribution, together with the non-Gaussian $f_{E^{\prime(\prime)}|S_{E^{\prime(\prime)}}}$, then allows for calculating the posterior $f_{S_{E^{\prime(\prime)}}|E^{\prime(\prime)}}$ that serves as the ``prior'' pdf for the next iteration. Explicitly, denoting the $i$th iteration in this scheme by the superscript `$(i)$', the iteration process is specified by
\bea
f^{(i+1)}_{S_{E^{\prime(\prime)}}}
=
f^{(i)}_{S_{E^{\prime(\prime)}}|E^{\prime(\prime)}} 
\label{eq:Bayes_iter}
\eea
for $i=0,1,\ldots$, thus yielding $f^{(n)}_{E^{\prime(\prime)}|S_{E^{\prime(\prime)}}}$ after $n$ iterations, with similar relations for the intensity and amplitude.

In general, (\ref{eq:Bayes})--(\ref{eq:Bayes_iter}) can be used even with empirical pdfs $f_{E^{\prime(\prime)}|S_{E^{\prime(\prime)}}}$.
A similar equation and procedure can be used for determining the intensity or magnitude for incoherent detection.
This outline of a procedure is sketched here for sake of completeness and guidance. However, no explicit results are included because they require knowledge of $f_{E^{\prime(\prime)}|S_{E^{\prime(\prime)}}}$ which must follow from a separate investigation, e.g., from experiment. Further study is needed with respect to the demonstration of convergence of (\ref{eq:Bayes_iter}) to a stable pdf.

\section{Confidence intervals\label{sec:confint}}
\begin{figure}[htb] \begin{center} \begin{tabular}{c} \ 
\epsfxsize=7cm \epsfbox{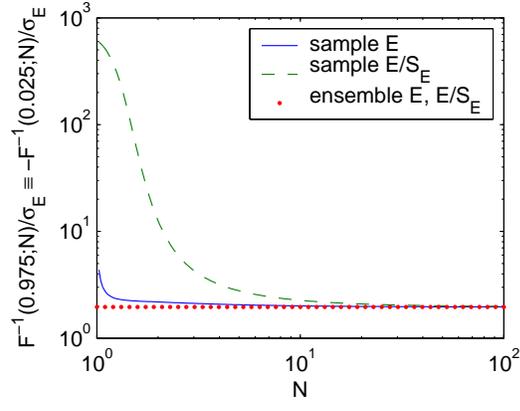}\  \\
\end{tabular} \end{center}
\caption{\label{fig:confintE}
\small
(color on-line)
Upper boundaries of $95\%$-confidence intervals
for real or imaginary parts of 
$E_\alpha$, $E_{\rm t}$, or $E$.
The lower boundary $F^{-1}(0.025) \equiv - F^{-1}(0.975)$ is symmetric with respect to $e=0$.
}
\end{figure}

\begin{figure}[htb] \begin{center} \begin{tabular}{c} \ 
\epsfxsize=7.5cm \epsfbox{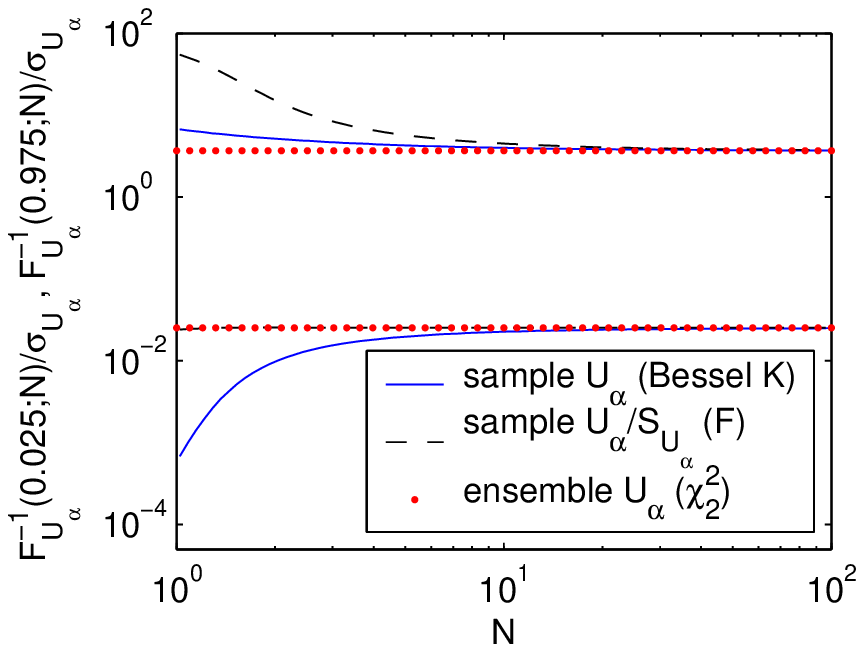}\ \\
(a) \\
\epsfxsize=7.5cm \epsfbox{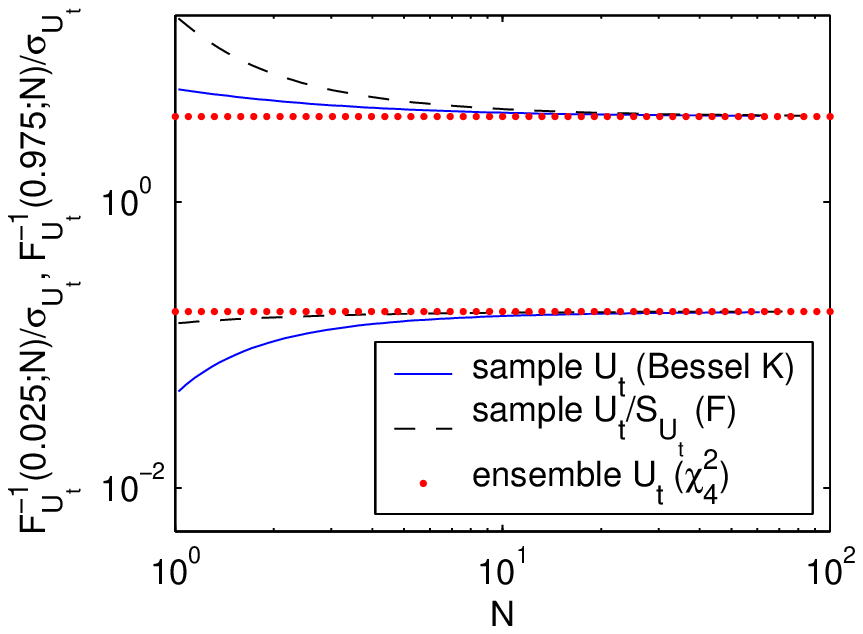}\ \\
(b) \\
\epsfxsize=7.5cm \epsfbox{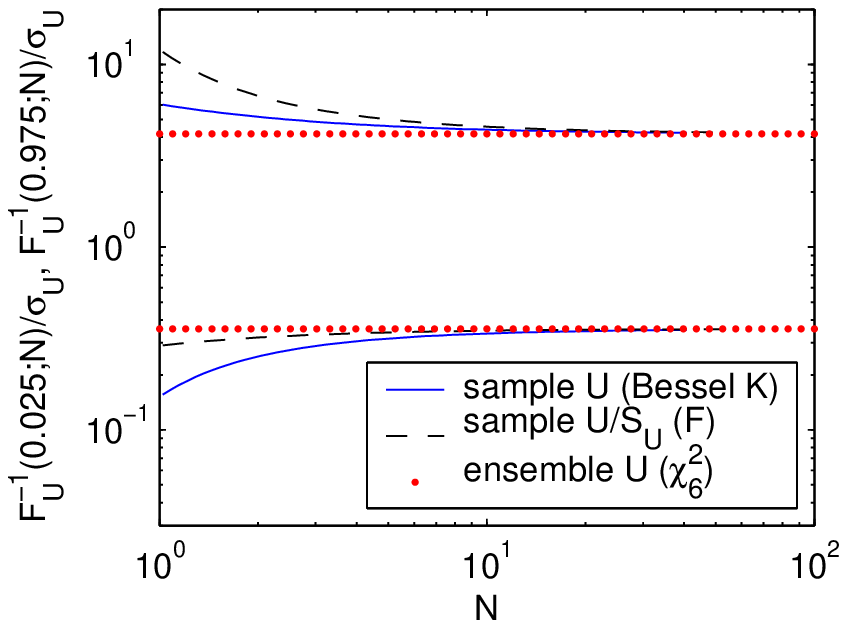}\ \\
(c)
\end{tabular} \end{center}
\caption{\label{fig:confintP}
\small
(color on-line)
Lower and upper boundaries of $95\%$-confidence intervals
(a) for a 1D Cartesian component $U_\alpha$ ($p{=}1$),
(b) for a 2D planar field $U_{\rm t}$ ($p{=}2$), and 
(c) for the 3D vectorial $U$ ($p{=}3$), normalized by the respective sampling standard deviations.
For each line type, the lower and upper curves represent $2.5\%$ and $97.5\%$ percentiles, respectively.
}
\end{figure}

\begin{figure}[htb] \begin{center} \begin{tabular}{c} \ 
\epsfxsize=7.5cm \epsfbox{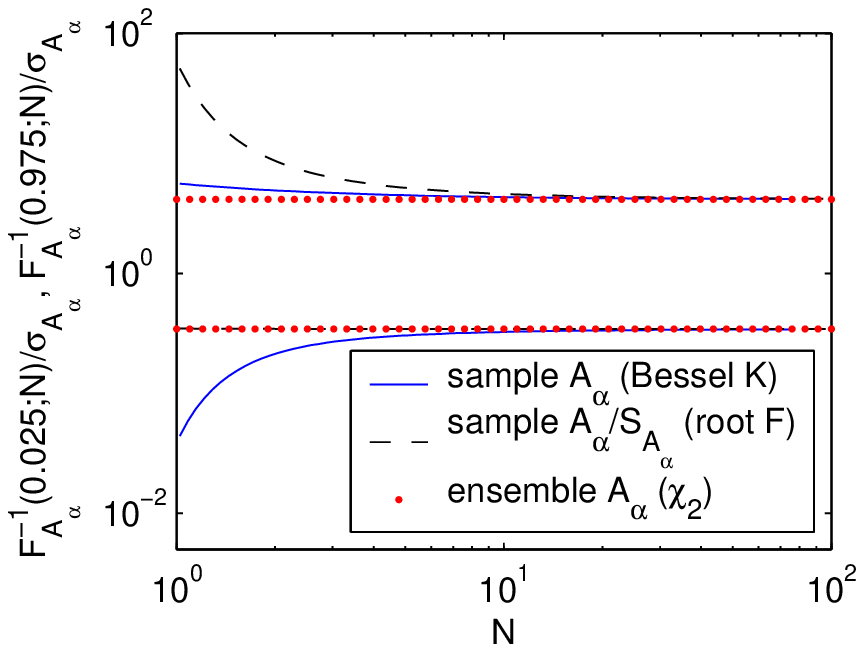}\ \\
(a) \\
\epsfxsize=7.5cm \epsfbox{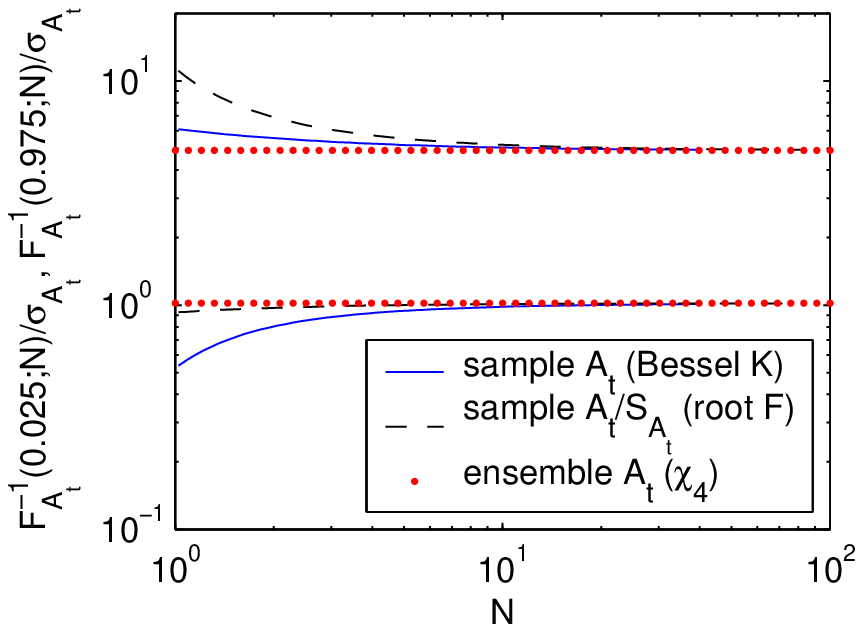}\ \\
(b) \\
\epsfxsize=7.5cm \epsfbox{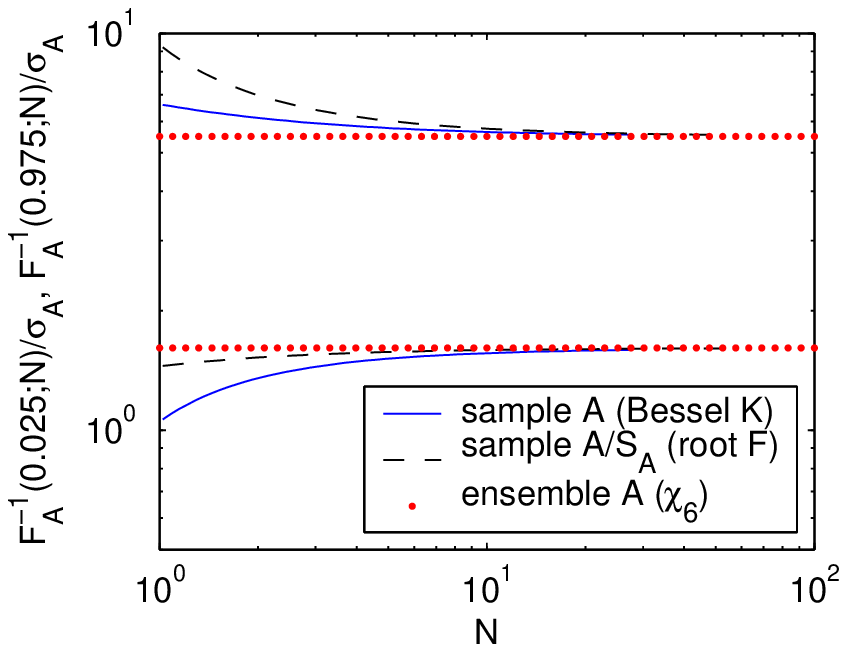}\ \\
(c)
\end{tabular} \end{center}
\caption{\label{fig:confintA}
\small
(color on-line)
Lower and upper boundaries of $95\%$-confidence intervals
(a) for a 1D Cartesian component $A_\alpha$ ($p{=}1$),
(b) for a 2D planar field $A_{\rm t}$ ($p{=}2$), and 
(c) for the 3D vectorial $A$  ($p{=}3$), normalized by the respective sampling standard deviations.
For each line type, the lower and upper curves represent $2.5\%$ and $97.5\%$ percentiles, respectively.
}
\end{figure}

$95\%$-confidence interval boundaries from sampling pdfs of $E$, $U$, $A$ as well as $X$, $W$, $V$ and their components, are compared to corresponding boundaries for the ensemble pdfs in Figs. \ref{fig:confintE}, \ref{fig:confintP} and \ref{fig:confintA}, respectively.
The Figures show that $N \sim 10$ or larger is needed in order for the sampling confidence interval boundaries to be close to the ensemble boundaries. For example, for an overmoded Fabry-P\'{e}rot resonator, a resonator length of the order of five wavelengths or larger is needed to achieve this.

\section{Conclusion\label{sec:conclusion}}
In this paper, we studied sampling distributions for the analytic complex-valued field, the intensity (energy density, power) and the magnitude for Gaussian statistically homogeneous random electromagnetic waves. 
The main results are 
(\ref{eq:samplepdfE_coh_BesselK})--(\ref{eq:samplepdfE_coh_BesselK_determbias}), 
(\ref{eq:samplepdfE_coh_BesselK_randombias}),
(\ref{eq:samplepdfX_coh_StudentT}),
(\ref{eq:samplepdfU_incoh_BesselK}), 
(\ref{eq:samplepdfW_incoh_FisherF}), 
(\ref{eq:samplepdfA_incoh_BesselK}) and 
(\ref{eq:samplepdfV_final_copy}).
A common feature is that lowering the number of degrees of freedom characterizing sampling distributions results in their tails becoming heavier and a consequent associated increase of widths of confidence intervals compared to the Gauss normal, $\chi^2_{2p}$ and $\chi_{2p}$ ensemble distributions, respectively. 
Furthermore, it was shown that standardized quantities (i.e., the sampled random field, magnitude or intensity divided by its own sample standard deviation, whereby the latter is itself considered as a random variable whose sample value is calculated from the sample data set itself) exhibit sampling pdfs that are characterized by a wider spread than for the case where that quantity has an {\it a priori} known (ensemble) standard deviation. The reason is that in the former case the ratio of random variables defines a bivariate sampling distribution whose (univariate) marginal pdf is sought, whereas in the latter case the sampling pdf is univariate from the outset.
The differences in sampling distributions arising from choosing normalization rather than standardization were shown to be negligible in all practical cases.

While the focus in this paper was on the simplest of {\it a priori} chosen conditional pdfs for the local instantaneous field in unbounded media, extension to some more general cases and boundary-value problems is straightforward. For example, compound exponential distributions for anisotropic ideal random fields near a conducting or dielectric boundary \cite{arnaRS,arnaEMC} can be used instead, leading to the total (vectorial) intensity and amplitude of 3-D random fields near a planar isotropic surface by simple superposition.

\section{Acknowledgements}
This work was supported in part by the Physical Programme of the U.K. National Measurement System Policy Unit 2006--2009.
I wish to thank Dr. P. Harris (NPL) for comments to the manuscript, Mr. S. Burbridge (Imperial College) for assistance with SGI Altix high-performance computation relating to Fig. 
\ref{fig:pdfP_paramNt_coh_coh_Cart_BesselK_log}, 
and Dr. M. Little (University of Oxford, U.K.) for discussions relating to ref. \cite{sprin1}.

\clearpage

\appendix

\section{Sampling distributions of field, intensity and amplitude\label{app:samplepdfsUandA}}
\subsection{Field\label{app:samplepdfX}}
Here we derive the one-dimensional sampling distribution for the intensity $|E|^2$ or energy density $U\propto |E|^2$ of a random statistically homogeneous Cartesian or vector electric (or, by extension, magnetic) field $E$. 

The pdf of an ideal random field $E$ 
is circular with independent and identically distributed (i.i.d.) real (in-phase) and imaginary (quadrature) components  
$E^\prime\stackrel{\Delta}{=} {\rm Re}(E)$ and 
$E^{\prime\prime} \stackrel{\Delta}{=} {\rm Im}(E)$. As a result,
these components can be studied in isolation from each other. 
The parent (ensemble) distribution of $E$ is a central
Gauss normal distribution ($\langle E \rangle = 0$).
Therefore, the standardized $N$-point sample variance 
$D_{N-1} \stackrel{\Delta}{=} (N{-}1)\thinspace S^2_{E^{\prime(\prime)}}/\sigma^2_{E^{\prime(\prime)}} = \sum^{N}_{i=1} (E^{\prime(\prime)}_i{-}M_{E^{\prime(\prime)}})^2/\sigma^2_{E^{\prime(\prime)}}$ exhibits a $\chi^2_{N{-}1}$ pdf, while $D_1 \stackrel{\Delta}{=} (E^{\prime(\prime)}-M_{E^{\prime(\prime)}})^2 / \sigma^2_{E^{\prime(\prime)}}$ has a $\chi^2_1$ pdf. Consequently, the ratio
\bea
X^{\prime(\prime)} \stackrel{\Delta}{=} \frac{E^{\prime(\prime)}-M_{E^{\prime(\prime)}}}{S_{E^{\prime(\prime)}}} = \sqrt{\frac{D_1}{D_{N-1}/(N{-}1)}}
\label{def:Xpp}
\eea
has a Student $t$ sampling distribution with $N{-}1$ degrees of freedom, whence for $E^{\prime(\prime)}$ itself,
\bea
f_{E^{\prime(\prime)}}\left (e^{\prime(\prime)};N \right ) 
&=& 
\frac{\Gamma\left ( \frac{N}{2} \right )}{\sqrt{\pi} ~\Gamma\left (\frac{N-1}{2} \right )~ \sqrt{N-1} 
}\nonumber\\
&\times& \left [ 1 +
\frac{1}{N-1} \left ( \frac{e^{\prime(\prime)}-m_{E^{\prime(\prime)}} }{s_{E^{\prime(\prime)}}} \right )^2
\right ]^{-N/2}
\label{eq:tE}
\eea
for $N>1$,
where 
$e^{\prime(\prime)}$, $m_{E^{\prime(\prime)} }$ and
$s_{E^{\prime(\prime)}}$ are sample values. 
Compared to the sample value $s_{E^{\prime(\prime)}}$ of the parent pdf, the sampling pdf exhibits an increased sampling standard deviation, viz., $\sqrt{(N{-}1)/(N{-}3)} s_{E^{\prime(\prime)}}$ for $N>3$.

Typically, the Student $t$ distribution arises in the characterization of the sample mean, i.e., for $[(M_{E^{\prime(\prime)}}-\langle E^{\prime(\prime)} \rangle)/(\sigma_{E^{\prime(\prime)}}/\sqrt{N})]/(S_{E^{\prime(\prime)}}/\sigma_{E^{\prime(\prime)}})$, instead of $[({E^{\prime(\prime)}}-M_{E^{\prime(\prime)}})/\sigma_{E^{\prime(\prime)}}] / (S_{E^{\prime(\prime)}}/\sigma_{E^{\prime(\prime)}})$ as in (\ref{def:Xpp}). Without lack of generality, however, we further use the simplified definition $X^{\prime(\prime)} \stackrel{\Delta}{=} {E^{\prime(\prime)}}/{S_{E^{\prime(\prime)}}}$, i.e., $M_{E^{\prime(\prime)}}=0$, because each sample set can always be centralized by its own sample mean value while maintaining its sampling pdf.

The square of $X^{\prime(\prime)}$, represented as $Y^{\prime(\prime)}/S_{Y^{\prime(\prime)}}\stackrel{\Delta}{=} X^{{\prime(\prime)}^2} = {E^{\prime(\prime)}}^2/S^2_{E^{\prime(\prime)}}$, exhibits a $t^2$ sampling distribution that follows 
from 
(\ref{eq:tE}) 
as 
\bea
f_{Y^{\prime(\prime)}}(y^{\prime(\prime)};N) &=& \frac{\Gamma\left ( \frac{N}{2} \right )}{2\sqrt{\pi} ~\Gamma\left (\frac{N-1}{2} \right )~\sqrt{N-1} }\nonumber\\
&~&\times
\left [ \frac{y^{\prime(\prime)}}{s_{Y^{\prime(\prime)}}} \left ( 1 + \frac{y^{\prime(\prime)}}{(N-1)\thinspace s_{Y^{\prime(\prime)}}}\right )^{N} \right ]^{-1/2}.
\label{eq:pdfY2}
\eea
which is a Fisher-Snedecor $F_{1,N-1}$ pdf of the ratio of the standard $\chi^2_1$ variate $D_1$ and the standard $\chi^2_{N-1}$ variate $D_{N-1}$ (both being statistically independent, because $M_{E^{\prime(\prime)}}$ and $S_{E^{\prime(\prime)}}$ are independent for Gauss normal $E^{\prime(\prime)}$), whereby each variate is divided by its corresponding number of degrees of freedom, i.e.,
\bea
t^2 = F_{1,N-1} = \frac{\chi^2_1/1}{\chi^2_{N-1}/(N-1)}.
\eea

\subsection{Field intensity\label{app:samplepdfW}}
To find the sampling distribution of a squared Cartesian or vectorial EM field with $2p$ i.i.d. components $Y^{\prime(\prime)}_i$ (corresponding to a $p$-dimensional analytic complex-valued field), i.e., $\sigma_{Y^{\prime(\prime)}_i}=\sigma^2_X$, we consider the ratio $Z/D_Z$, where
\bea
Z \stackrel{\Delta}{=} \frac{Y^\prime+Y^{\prime\prime}}{\sigma_{Y^{\prime(\prime)}}} 
=
\sum^{p}_{i=1} \left ( \frac{Y^\prime_i}{\sigma_{Y^\prime}} + \frac{Y^{\prime\prime}_i}{\sigma_{Y^{\prime\prime}}} \right ) 
\sum^{2p}_{i=1} \frac{(X_i-m_X)^2}{\sigma^2_X} ~~~
\eea 
has a standard $\chi^2_{2p}$ distribution, on account of the addition theorem for $2p$ i.i.d. standard $\chi^2_1$ variates, and 
\bea
D_Z\stackrel{\Delta}{=}\frac{(2pN-1) ~ S^2_{X^{\prime(\prime)}}}{\sigma^2_{X^{\prime(\prime)}}}
\eea
exhibits a standard $\chi^2_{2pN{-}1}$ distribution, with $S^2_{X^{\prime(\prime)}} = $\linebreak $\sum^{2pN}_{i=1} (X^{\prime(\prime)}_i{-}M_{X^{\prime(\prime)}})^2/(2pN{-}1)$. 
Hence, their ratio $G$ is a scaled Fisher-Snedecor $F$ variate with ($2p,2pN-1)$ degrees of freedom:
\bea
G 
\stackrel{\Delta}{=}
\frac{Z}{D_Z}
&\equiv&
\frac{2p}{2pN-1} \frac{Z/(2p)}{D_Z / (2pN-1)}\nonumber\\
&=&
\frac{2p}{2pN-1} F_{2p,2pN-1}.
\label{eq:ZKZ_1}
\eea
With $\sigma_{|E|^2}=2\sqrt{p}\thinspace \sigma^2_{E^{\prime(\prime)}}$, the ratio $Z/D_Z$ can be related to the standardized power or field intensity as
\bea
\frac{U}{S_{U}}
= \frac{\sum^{2p}_{i=1} {E^\prime_{i}}^2}{2\sqrt{p} ~ S^2_{E^\prime}}
&=& \frac{2pN-1}{2\sqrt{p} } \frac{\sum^{2p}_{i=1} (E^\prime_{i} / \sigma_{E^\prime})^2}{(2pN-1)~ S^2_{E^\prime} /\sigma^2_{E^\prime}}\nonumber\\
&=& \frac{2pN-1}{2\sqrt{p} } \frac{Z}{D_Z}.
\label{eq:ZKZ_2}
\eea
Thus, combining (\ref{eq:ZKZ_1}) and (\ref{eq:ZKZ_2}) yields
\bea
W \stackrel{\Delta}{=}
\frac{U}{S_U} 
=
\sqrt{p} ~ F_{2p,2pN-1}.
\label{eq:ZKZ_3}
\eea
With (\ref{eq:ZKZ_1}), upon scaling the standard $F_{2p,2pN-1}$ distribution, the sampling pdf of $G$ is
\bea
f_G(g;N) 
&=& \frac{\Gamma\left (pN-\frac{1}{2}+p \right )}{\Gamma(pN-\frac{1}{2})\Gamma(p)}~
\frac{g^{p-1}}{\left ( 1 + g \right )^{pN-\frac{1}{2}+p}}~~~
.
\label{eq:pdfZ_general}
\eea
For the square of the Cartesian in-phase field component $[{\rm Re}(E)]^2$ or the quadrature component $[{\rm Im}(E)]^2$ (i.e., for $p=1/2$), (\ref{eq:pdfZ_general}) reduces to (\ref{eq:pdfY2}) as expected.
Finally, from (\ref{eq:ZKZ_3}), the sampling distribution of $W=U/S_{U}$ follows as 
\bea
f_W(w;N) 
&=& \frac{\Gamma\left ( pN-\frac{1}{2}+p \right )}{\Gamma(pN-\frac{1}{2})\Gamma(p)}~\frac{p^{p/2} }{\left ( pN-\frac{1}{2}\right )^{p}}\nonumber\\
&~&\times \frac{w^{p-1}}{\left ( 1 + \frac{\sqrt{p}}{pN-\frac{1}{2}} \thinspace w \right )^{pN-\frac{1}{2}+p}}.
\label{eq:samplepdfW_final}
\eea
This represents the pdf of the standardized received sampled power $W$ at $w$, i.e., for the ratio of sampled values $u$ and $s_{U}$. 

In the limit $N\rightarrow+\infty$, the exponent $p(N+1)-\frac{1}{2}$ in (\ref{eq:samplepdfW_final}) reduces to $pN-\frac{1}{2}$; the prefactor $\Gamma(pN-\frac{1}{2}+p)/\Gamma(pN-\frac{1}{2}) = (pN-\frac{1}{2})\cdot \ldots \cdot (pN-\frac{3}{2}+p)$ for $p\geq 1$ becomes $(pN-\frac{1}{2})^p$; and $s_{U}$ approaches the ensemble statistic $\sigma_{U}$, whence
\bea
f_{W}(w;N) \rightarrow f_U(u) = \frac{p^{p/2}}{\Gamma(p) \sigma_{U}}
\left ( \frac{u}{\sigma_{U}} \right )^{p-1} \exp \left ( - \sqrt{p} \thinspace \frac{u}{\sigma_{U}} \right )
\nonumber\\
\eea
which is the ensemble $\chi^2_{2p}$ limit pdf, as expected. The result agrees with a well-known limit theorem from probability theory  \cite{cram1} stating that $F_{m,n}(x)$ converges 
to $\chi^2_m(mx)$ when $n/m{\rightarrow}+\infty$, i.e., $F_q(m,n)\rightarrow \chi^2_q(m)/m$ for the corresponding quantiles. 
In the same limit $N{\rightarrow}{+}\infty$, the asymptotic mean value of $W$ is  
\bea
\langle W \rangle = \frac{2pN{-}1}{2pN{-}3} \sqrt{p} \rightarrow \sqrt{p}
\eea 
and the standard deviation for $pN > 5/2$ is
\bea
\sigma_W = \sqrt{\frac{2(2pN{-}1)^2(2pN{+}2p{-}3)}{2p(2pN{-}3)^2 (2pN{-}5)} {p}} \rightarrow 1.
\eea 
Thus, we find that the coefficient of variation, i.e.,
\bea
\frac{\sigma_{W}}{\langle {W}\rangle} = \sqrt{\frac{2pN{+}2p{-}3}{p(2pN{-}5)}} ~ \rightarrow \frac{1}{\sqrt{p}}
\eea
has its value reduced to that for the $\chi^2_{2p}$ ensemble distribution when $N{\rightarrow}+\infty$, whereas for small $N$ its value is substantially larger, indicating larger relative uncertainty.

\subsection{Field amplitude\label{app:samplepdfV}}
For the standardized field magnitude $V \stackrel{\Delta}{=} A/S_{A} =\sqrt{(S^2_V/S_W) ~W}$, with 
\bea
\frac{\sigma^2_Z}{\sigma_W} = \frac{\sigma^2_{A}}{\sigma_{A^2}} =
\frac{p - \left ( \frac{\Gamma\left (p+\frac{1}{2} \right )}{\Gamma(p)} \right )^2}{\sqrt{p}}
\eea
valid for $\chi^{(2)}_{2p}$ distributions, from variate transformation of (\ref{eq:samplepdfW_final}) and noting that
\bea
\frac{w}{s_W} = \left ( \frac{v}{s_V} \right )^2 = \frac{v^2}{s_W} \cdot \frac{s_W}{s^2_V},~~
{\rm d}w = 2 v \frac{s_W}{s^2_V} {\rm d}v,~~~
\eea
we obtain the sampling pdf of $V$ at $v=a/s_{A}$ as
\bea
f_V(v;N) 
&=& C_V
{v^{2p-1}}\nonumber\\
&~& \times {\left [ 1 + \frac{p-\left ( \frac{\Gamma\left ( p+\frac{1}{2} \right )}{\Gamma(p)}\right )^2}{pN-\frac{1}{2}} \thinspace v^2 \right ]^{-(pN+p-\frac{1}{2})}}
\label{eq:samplepdfV_incoh_rootF}
\eea
which can be referred to as a root-$F$ distribution, where
\bea
C_V \stackrel{\Delta}{=} \frac{2}{\left (pN-\frac{1}{2}\right )^{p}}
\left [ p - \left ( \frac{\Gamma\left ( p+\frac{1}{2} \right )}{\Gamma(p)}\right )^2 \right ]^{p} \frac{\Gamma\left (pN-\frac{1}{2}+p \right )}{\Gamma(pN-\frac{1}{2})\Gamma(p)}. \nonumber\\
\eea

\section*{Figures}

\clearpage
\clearpage
\clearpage
\clearpage
\clearpage
\clearpage
\clearpage
\clearpage
\clearpage
\clearpage
\clearpage
\clearpage
\clearpage
\clearpage

\end{document}